\newif\ifconfver
\confverfalse      

\newif\ifcutshort      
\cutshorttrue

\newif\ifcutshortlvltwo  
\cutshortlvltwofalse

\ifconfver
\documentclass[10pt,twocolumn,twoside]{IEEEtran}
\else
\documentclass[11pt,draftcls,onecolumn]{IEEEtran}
\fi

\usepackage{array}
\usepackage{cite}
\usepackage{calc}
\usepackage{caption}
\usepackage{graphicx}
\usepackage{psfrag}
\usepackage[tight,footnotesize]{subfigure}
\usepackage{stfloats}
\usepackage{url}
\usepackage{subfig}
\usepackage{longtable}
\usepackage{amsmath,amsfonts,amssymb,amsbsy,bm,paralist,theorem,ifthen,color}
\usepackage{algorithmic,algorithm,theorem}
\usepackage[dvips]{epsfig}
\usepackage[dvips]{graphicx}

\newcommand\Fc{\ensuremath{\mathcal{F}}}

\newcommand\Lc{\ensuremath{{\mathcal{L}}}}

\newcommand\Kc{\ensuremath{{\mathcal{K}}}}

\newcommand\xb{\ensuremath{{\bm x}}}

\newcommand\wb{\ensuremath{{\bm w}}}
\newcommand\yb{\ensuremath{{\bm y}}}

\newcommand\ub{\ensuremath{{\bm u}}}
\newcommand\Hb{\ensuremath{{\bm H}}}
\newcommand\hb{\ensuremath{{\bm h}}}
\newcommand\Ab{\ensuremath{{\bm A}}}
\newcommand\ab{\ensuremath{{\bm a}}}
\newcommand\Bb{\ensuremath{{\bm B}}}
\newcommand\bb{\ensuremath{{\bm b}}}
\newcommand\Cb{\ensuremath{{\bm C}}}

\newcommand\eb{\ensuremath{{\bm e}}}

\newcommand\Gb{\ensuremath{{\bm G}}}
\newcommand\gb{\ensuremath{{\bm g}}}
\newcommand\Ib{\ensuremath{{\bm I}}}
\newcommand\Ic{\ensuremath{{\mathcal{I}}}}

\newcommand\pb{\ensuremath{{\bm p}}}
\newcommand\Sb{\ensuremath{{\bm S}}}
\newcommand\Tb{\ensuremath{{\bm T}}}

\newcommand\Rb{\ensuremath{{\bm R}}}
\newcommand\Mb{\ensuremath{{\bf M}}}

\newcommand\Qb{\ensuremath{{\bm Q}}}

\newcommand\vb{\ensuremath{{\bm v}}}
\newcommand\Wb{\ensuremath{{\bm W}}}

\newcommand\zb{\ensuremath{{\bm z}}}

\newcommand\mub{\ensuremath{{\bm \mu}}}
\newcommand\lambdab{\ensuremath{{\bm \lambda}}}
\newcommand\nub{\ensuremath{{\bm \nu}}}
\newcommand\Omegab{\ensuremath{{\bm \Omega}}}
\newcommand\Upsilonb{\ensuremath{{\bm \Upsilon}}}
\newcommand\zerob{\ensuremath{{\bm 0}}}

\newcommand\LambdaB{\ensuremath{{\bf \Lambda}}}

\newcommand\Phib{\ensuremath{{\bf \Phi}}}

\newcommand\SigmaB{\ensuremath{{\bf \Sigma}}}
\newcommand\sigmab{\ensuremath{{\bm \sigma}}}

\newcommand\E{\ensuremath{{\mathbb{E}}}}

\newcommand\diag{\ensuremath{{\rm diag}}}

\newcommand\oneb{\ensuremath{{\bf 1}}}

\newtheorem{Lemma}{Lemma}
\newtheorem{Proposition}{Proposition}
\newtheorem{Theorem}{Theorem}

\ifconfver

\else

\fi

\begin{document}

    \bibliographystyle{IEEEtran}

    \title{QoS-Based Linear Transceiver Optimization for Full-Duplex Multi-User Communications}

    \ifconfver \else {\linespread{1.1} \rm \fi

        \author{\vspace{0.8cm}Tsung-Hui Chang, Ya-Feng Liu and Shih-Chun Lin\\
            \thanks{
    T.-H. Chang is the corresponding author. He is with the School of Science and Engineering, The Chinese University of Hong Kong, Shenzhen, Shenzhen, China (e-mail: tsunghui.chang@ieee.org)}
    \thanks{Y.-F.~Liu is with the State Key Laboratory
    of Scientific and Engineering Computing, Institute of Computational
    Mathematics and Scientific/Engineering Computing, Academy of
    Mathematics and Systems Science, Chinese Academy of Sciences,
    Beijing, 100190, China (e-mail:
    {{yafliu}@lsec.cc.ac.cn}).}
    \thanks{S.-C. Lin is with the Department of Electronic and Computer Engineering, National Taiwan University of Science and
    	Technology, Taipei, Taiwan (email: sclin@mail.ntust.edu.tw).}
    \thanks{The work of T.-H. Chang is supported by NSFC, China, Grant No. 61571385. 
    	The work of Y.-F. Liu is supported by the NSFC, China,
    Grant No. 11671419 and 11631013.
    The work of S.-C. Lin is supported by MOST, Taiwan, Grant No. MOST 104-2628-E-011-008-MY3.} 
    \thanks{Part of this work is presented in IEEE GLOBECOM 2016 \cite{CLOBECOM16}.}
            }

        \maketitle

        \begin{abstract}
           In this paper, we consider a multi-user wireless system with one full duplex (FD) base station (BS) serving a set of half duplex (HD) mobile users. 
           To cope with the in-band self-interference (SI)
           and co-channel interference, we formulate a quality-of-service (QoS) based linear transceiver design problem. 
           The problem jointly optimizes the downlink (DL) and uplink (UL) beamforming vectors of the BS and the transmission powers of UL users so as to provide both the DL and UL users with guaranteed signal-to-interference-plus-noise ratio performance, using a minimum UL and DL transmission sum power. 
           The considered system model not only takes into account noise caused by non-ideal RF circuits, analog/digital SI cancellation but also constrains the maximum signal power at the input of the analog-to-digital converter (ADC) for avoiding signal distortion due to finite ADC precision. The formulated design problem is not convex and challenging to solve in general. We first show that for a special case where the SI channel estimation errors are independent and identically distributed, the QoS-based linear transceiver design problem is globally solvable by a polynomial-time bisection algorithm.
           For the general case, we propose a suboptimal algorithm based on alternating optimization (AO). The AO algorithm is guaranteed to converge to a Karush-Kuhn-Tucker solution.
           To reduce the complexity of the AO algorithm, we further develop a fixed-point method by extending the classical uplink-downlink duality in HD systems to the FD system. 
           Simulation results are presented to demonstrate the performance of the proposed algorithms and the comparison with HD systems.
            \\\\
            \noindent {\bfseries EDICS}:  SAM-BEAM, SPC-INTF, SPC-APPL, SAM-APPL.
        \end{abstract}

        \ifconfver \else
        \newpage
        \fi

        \ifconfver \else \IEEEpeerreviewmaketitle} \fi

    \vspace{-0.3cm}

\section{Introduction}\label{sec: intro}

The next generation wireless communication systems target at ten times faster transmission rates and much shorter latency than the current 4G system. In addition to more powerful coding schemes and advanced multiple antenna techniques, the full duplex (FD) technique has also been considered as a solution with great potential to reach the target \cite{PitavalWC15}.  
Ideally, a FD system can double the spectral efficiency compared to the conventional half duplex (HD) systems since it allows the node to transmit and receive signals at the same time and over the same frequency  \cite{IBFDCO}.
In practice, however, simultaneous transmission and reception cause severe self-interference (SI) which might greatly limit the system performance. Fortunately, recent advances in analog and
digital SI cancellation (SIC) techniques have made the FD technique
successfully implemented in bi-directional
\cite{Day2012bidirectional}, relay \cite{Day2012relay} and WiFi
\cite{EDCOFDWS,Duarte2014,Katti_14} systems.

Recently, there have been of great interest to consider the FD
techniques in the multi-user cellular systems \cite{FDCSWDIPDC}.
Specifically, in such scenarios, a FD base station (BS) can serve both the downlink mobile users (DMUs) and uplink mobile users (UMUs) simultaneously. However, new challenges arise as
not only the BS suffers from the SI, but also the DMUs are interfered by UMUs. This new form of uplink-to-downlink (UL-to-DL) co-channel interference (CCI) could become the performance bottleneck if not appropriately mitigated. In fact, the SI and UL-to-DL CCI couples the DL and UL transmissions, and therefore unlike the HD systems where the UL and DL transmissions can be designed separately, the FD system must consider a joint UL and DL transmission design.
There have been considerable efforts studying the FD joint design problems in the literature; see, e.g., 
\cite{CirikTC15,Nguyen2013,Nguyen2014,KimVT16,LiICC14,ChoiTWC2014,CirikWCL16,SunTWC16}.
Most of the works have focused on the multiple-input multiple-output (MIMO) scenarios by assuming that both the BS and MUs are equipped with multiple antennas. For example, \cite{CirikWCL16,Nguyen2013,Nguyen2014,KimVT16,LiICC14} have studied beamforming/precoding and resource allocation algorithms for network sum rate or energy efficiency maximization, while \cite{CirikTC15,SunTWC16} have 
considered algorithms for providing the MUs with guaranteed quality-of-service (QoS).

In this paper, we consider a multiple-input single-output (MISO) scenario where each of the MUs has a single antenna. The multi-antenna FD BS employs transmit beamforming for DL transmission \cite{Gershman2010_SPM}.
Unlike \cite{CirikWCL16,Nguyen2013,Nguyen2014,KimVT16,LiICC14}, which assumed the optimal non-linear receiver for UL signal detection, we consider a linear receive beamforming scheme for signal-user detection at the BS \cite{Gershman2010_SPM}.
Under these settings, we formulate a QoS-based linear transceiver design problem that minimizes the sum of UL and DL transmission powers subject to constraints that guarantee minimum signal-to-interference-plus-noise ratio (SINR)  requirement of MUs.
Like the MIMO formulations considered in \cite{CirikWCL16,Nguyen2013,Nguyen2014,KimVT16,LiICC14}, the MISO QoS-based linear transceiver problem is still difficult to solve \cite{SunTWC16} due to the coupled DL and UL transmissions. We are interested in such problem formulation for two reasons. First, the QoS based formulation is more fundamental as the solution to the problem can be  used for other design formulations such as the max-min-fairness designs \cite{Wiesel2006} and rate region characterization. Second, in HD systems, the DL QoS-based beamforming problem \cite{BK:Bengtsson01} and UL QoS-based receive beamforming and power control problem \cite{LiuHongDaiSPL13} are known polynomial-time solvable, though they are not convex problems in their original forms.
Specifically, there exists an uplink-downlink duality (UDD) \cite{Liu1998,Sch04,Wiesel2006} between the UL and DL problems and the two problems can be efficiently solved by a fixed-point iterative method \cite{Wiesel2006,Liu1998}. Therefore, it is interesting to see whether these elegant results can be generalized to the FD systems.
The main contributions of this paper are threefold:
\begin{itemize}
	\item {\bf Practical problem formulation:} As motivated by the circuit designs in \cite{Katti_ACM13,Katti_14,Duarte2014},
	we consider a system model that not only takes into account the noise caused by non-ideal RF chains but also the capability of analog and digital SIC schemes.
	The proposed formulation based on this practical model thus captures the impact of these relevant system parameters on the network performance. In particular, different from the existing works \cite{CirikTC15,Nguyen2013,Nguyen2014,KimVT16,LiICC14,ChoiTWC2014,CirikWCL16,SunTWC16}, the formulated QoS-based design problem explicitly constrains the signal power level at the input of analog-to-digital converter (ADC), for avoiding signal distortion due to limited ADC precision. Such constraint is critical to the design of a FD transceiver \cite{Katti_ACM13}, but has not been explicitly considered in the literature.
	
	\item {\bf Polynomial-time solvable subclass:} While the QoS-based linear transceiver problem is non-convex and difficult to solve in general, we identify one intriguing case of the problem that is globally solvable.
	Specifically, we show that if the SI channel estimation error (i.e., the residual SI channel after analog SIC) has independent and identically distributed (i.i.d.) elements, then the non-convex problem can actually be globally solved by a bisection algorithm. The algorithm itself reveals interesting insights into the solution structure of the considered problem.
	
	\item  {\bf Efficient suboptimal algorithm:} In particular, our analysis suggests that it is unlikely to solve the QoS-based linear transceiver design problem in a convex fashion with respect to all the variables of the DL beamformer, UL beamformer and UMUs' transmission powers. In light of this, to handle the considered design problem in general, we propose an alternating optimization (AO) based suboptimal algorithm, which iteratively optimizes the UL beamformer followed by optimizing the DL beamformer and UMUs' transmission powers until convergence. Moreover, we generalize the UDD in HD systems  \cite{Liu1998,Sch04,Wiesel2006} to the FD system and developed a new fixed-point algorithm to improve the computational efficiency of the proposed AO algorithm.
\end{itemize}
Simulation results are presented to illustrate the performance of the proposed algorithms under various settings and comparison with the HD system.

{\bf Synopsis:} Section \ref{sec: signal model} presents the FD system model and the considered QoS-based design problem. 
Some existing results in HD systems are also briefly reviewed. Section \ref{sec: bisection} considers a special case of the considered problem and 
proposed a globally optimal bisection algorithm. In Section \ref{sec: AO}, an AO algorithm is proposed, and the UDD and fixed-point algorithm for the FD systems are presented.
Simulation results are given in Section \ref{sec: simulation} and conclusions are drawn in Section \ref{sec: conclusion}.

{\bf Notations:} $\diag(\Ab)$ is a diagonal matrix with its diagonal elements equal to the diagonal elements of matrix $\Ab$;  $\diag(\{a_i\}_i)$ and $\diag(\ab)$, where $\ab=[a_1,\ldots,a_m]^T$, both represent a diagonal matrix with $a_i$'s being the diagonal elements; we also denote
$|\ab|^2 = [|a_1|^2, \ldots, |a_m|^2]^T$ ;  $\eb_n$ denotes the elementary vector with one in the $n$th entry and zero otherwise.
$\ab \succeq \bb$ denotes element-wise inequality for vectors $\ab$ and $\bb$, while $\Ab\succeq (\succ) \Bb$  means that $\Ab-\Bb$ is a positive semidefinite (positive definite) matrix. $\lambda_{\max}(\Ab)$ denotes the maximum eigenvalue of matrix $\Ab$. ${\rm vec}(\Ab)$ is a vector obtained by stacking the columns of $\Ab$. Finally, $\Ib_{n}$ denotes the $n$ by $n$ identity matrix and $\E(\cdot)$ denotes the statistical expectation operator.

\section{Signal Model and Problem Formulation}\label{sec: signal model}


\subsection{FD Signal Model}\label{subsc: signal model}
\begin{figure}[!t]
	\begin{center}
		{\resizebox{.3\textwidth}{!}
			{\includegraphics{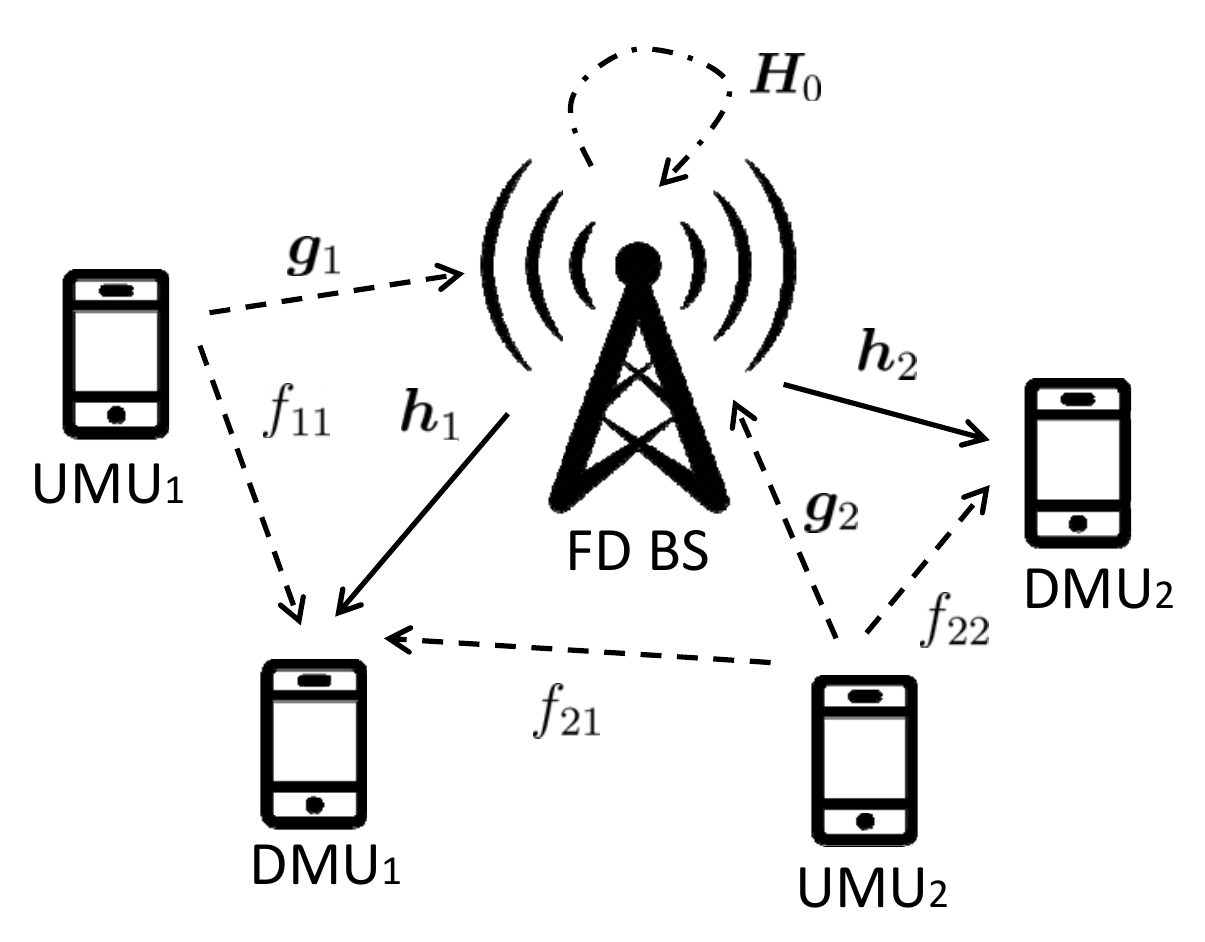}}}
	\end{center}\vspace{-0.3cm}
	\caption{A wireless network with one FD BS and multiple MUs.}
	\vspace{-0.3cm}\label{fig: network diagram}
\end{figure}

As shown in Figure \ref{fig: network diagram}, we consider a wireless system with one BS and a set of $K$ DMUs and $L$ UMUs. The DMUs want to receive information signal from the BS whereas the UMUs want to transmit information signals to the BS.
We assume that a FD BS,  which is  equipped with $N_t$ antennas ($N_t\geq 1$), is capable of communicating with the DMUs and UMLs
at the same time and over a common spectrum. 
The DMUs and UMUs are assumed to be HD and have a single antenna.

\begin{figure}[!t]
	\begin{center}
		{\resizebox{.35\textwidth}{!}
			{\includegraphics{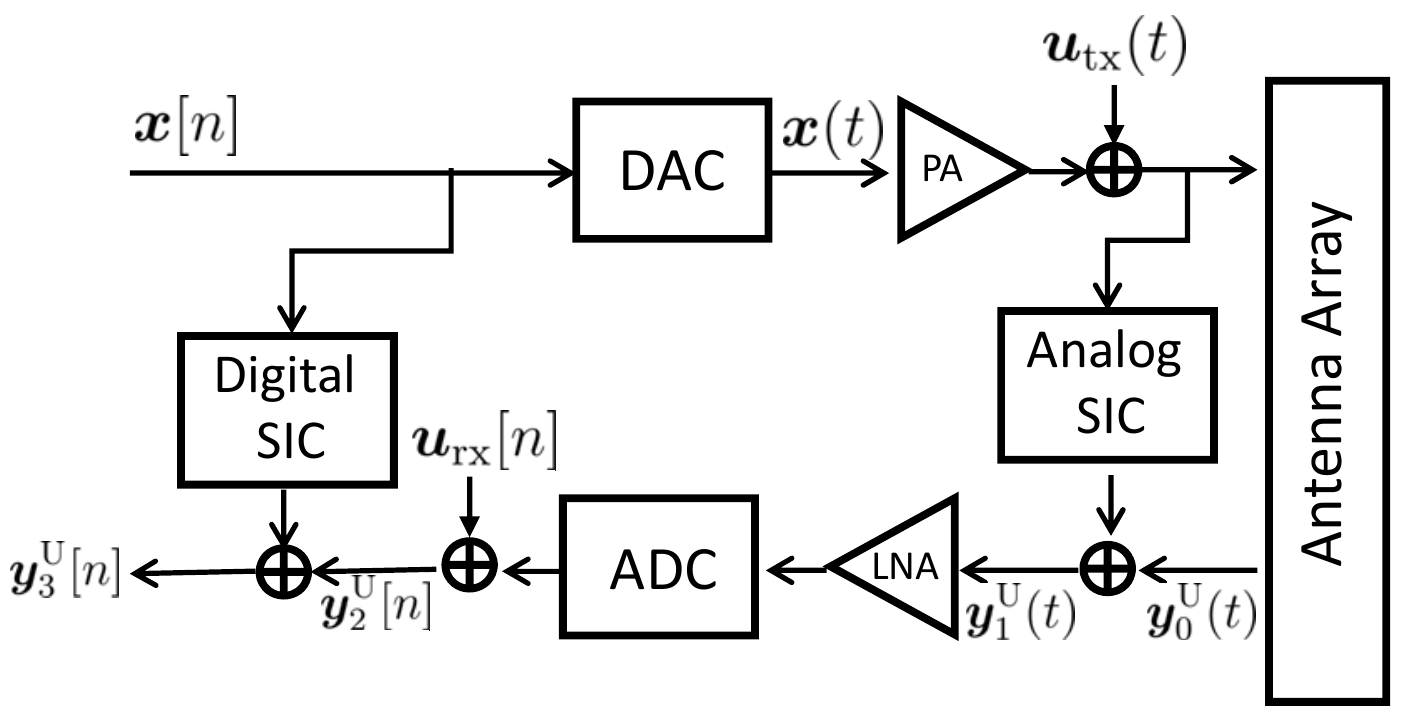}}}
	\end{center}\vspace{-0.3cm}
	\caption{A block diagram of the transmitter and receiver chains of the FD BS.}
	\vspace{-0.5cm}\label{fig: block diagram}
\end{figure}

We present a FD signal model which is motivated by the circuit designs proposed in \cite{Katti_ACM13,Katti_14} and the signal model in \cite{Day2012bidirectional}.
In particular, the proposed model explicitly accounts for the effects of analog SIC, digital SIC \cite{Katti_ACM13,Katti_14} as well as transmitter and receiver noises caused by non-linear circuit components \cite{Day2012bidirectional}, which therefore allows for detailed assessment of the FD system performance. 
The proposed model is illustrated in Figure \ref{fig: block diagram}.
Let $\xb[n]=\sum_{i=1}^K \wb_{i} s_i^{\rm D}[n]$ be the (discrete-time) signal transmitted by the BS to DMUs,
where $s_i^{\rm D}[n] \in \mathbb{C}$ is the (independent, zero mean and unit power)
information signal for DMU $i$, and $\wb_{i}\in \mathbb{C}^{{N_t}}$ is the associated beamforming vector, for all $i\in\Kc \triangleq \{1,\ldots,K\}$. As shown in Figure \ref{fig: block diagram}, due to non-ideal transmitter RF chain (e.g., non-linearity of digital-to-analog converter (DAC) and power amplifier), the continuous-time information signal $\xb(t) = \sum_{i=1}^K \wb_{i} s_i^{\rm D}(t) $ is corrupted by a transmitter noise signal $\ub_{\rm tx}(t) \in \mathbb{C}^{N_t}.$
Following \cite{Day2012bidirectional}, we model $\ub_{\rm tx}(t)$ as a complex Gaussian random process with zero mean and covariance matrix $\beta_1\diag(\sum_{i=1}^K \wb_i \wb_i^H)$ at each time $t$, i.e.,

\vspace{-0.2cm}
{\small  \begin{align}\label{eqn: u_tx}
\ub_{\rm tx}(t) \sim  \mathcal{CN}\bigg(\zerob,\beta_1\diag\bigg(\sum_{i=1}^K \wb_i \wb_i^H\bigg)\bigg),
\end{align}}
\!\!where $\beta_1 \ll 1$ is a constant. The noise  $\ub_{\rm tx}(t)$ is assumed to be independent of $\xb(t)$ and the receiver noise. The combined signal $\xb(t)+\ub_{\rm tx}(t)$ is then transmitted to the DMUs through the antenna array.

\subsubsection{Downlink user SINR}\label{subsec: downlink SINR}
Since the DMUs and UMUs communicate with the BS simultaneously, in addition to the signal from the BS, the DMUs also receive signals from the UMUs.
Let $\hb_i\in \mathbb{C}^{N_t}$ be the channel vector between the BS and the $i$th DMU, and let $f_{j i}\in \mathbb{C}$ be the channel coefficient from the UMU $j$ to DMU $i$ (see Figure \ref{fig: network diagram}). Moreover, denote
$p^{\rm U}_\ell>0$ as the transmission power of the $\ell$th UMU,
and $s_\ell^{\rm U}(t)\in \mathbb{C}$ as the (zero mean, unit power)
UL information signal, for all $\ell\in \Lc \triangleq \{1,\ldots,L\}$.
Then the received signal
of each DMU $i$ is given by
\begin{align}\label{eqn: DMU receive signal}
y_i^{\rm D}(t) &= \hb_{i}^H \bigg(\sum_{k=1}^K \wb_{k} s_k^{\rm D}(t) + \ub_{\rm tx}(t) \bigg) + \sum_{j=1}^L f_{ji} \sqrt{p^{\rm U}_j} s_j^{\rm U}(t) + z^{\rm D}_i(t),
~\forall i\in \Kc,
\end{align}
where $z^{\rm D}_i(t) \in \mathbb{C}$ is the additive white Gaussian noise (AWGN) following $\mathcal{CN}(0,\sigma_i^2)$.
By \eqref{eqn: DMU receive signal}, the SINR of DMU $i$ can be shown as
\begin{align}\label{eqn: downlink sinr}
    & {\rm SINR}_{i}^{\rm D}(\Wb,\pb^{\rm U}) =  \\
    &\frac{|\hb_{i}^H\wb_{i}|^2 }
    {\sum_{k\neq i}^K |\hb_i^H\wb_k|^2+ \beta_1 \hb_i^H  \diag(\sum_{k=1}^K \wb_k\wb_k^H) \hb_i +
          \hat  \sigma_i^2(\pb^{\rm U})} , \notag
\end{align} for all $i\in \Kc$, where $\Wb\triangleq [\wb_1,\ldots,\wb_K]$, $\pb^{\rm U}\triangleq [p_1^{\rm U},\ldots,p_L^{\rm U}]^T$ and
$\hat  \sigma_i^2(\pb^{\rm U}) \triangleq \sum_{j=1}^{L} p_j^{\rm U}|f_{ji}|^2 +
\sigma_i^2.$
Note that the second and third terms in the denominator of \eqref{eqn: downlink sinr} are due to the transmitter noise $\ub_{\rm tx}(t)$, the UL-to-DL CCI $\sum_{j=1}^L f_{ji} \sqrt{p^{\rm U}_j} s_j^{\rm U}(t)$ and the additive noise.
Let $\gamma^{\rm D}_i>0$ be the target SINR for DMU $i$. Then the SINR constraint for DMU $i$ can be written as
\begin{align}\label{eqn: downlink sinr2}
&{\rm SINR}_{i}^{\rm D}(\Wb,\pb^{\rm U}) \geq \gamma^{\rm D}_i \notag \\
\Longleftrightarrow ~~&
    \frac{|\hb_{i}^H\wb_{i}|^2/\rho_i^{\rm D}  }
{ \sum_{k=1}^K \wb_k^H \tilde \Hb_i \wb_k +
   \hat  \sigma_i^2(\pb^{\rm U})  }\geq 1,
\end{align}
where $\tilde \Hb_i \triangleq \hb_i\hb_i^H +  \beta_{1} \diag(|\hb_i|^2),$ for all $i\in \Kc$, and $\frac{1}{\rho_i^{\rm D} }\triangleq  1+1/\gamma_i^{\rm D}$.

\subsubsection{Uplink user SINR}\label{subsec: uplink SINR}
Denote $\Hb_0\in \mathbb{C}^{N_t \times N_t}$ as the SI channel matrix and $\gb_{\ell}\in \mathbb{C}^{N_t}$ as the channel vector
from the $\ell$th UMU to the BS  (see Figure \ref{fig: network diagram}).
The signal received by the BS thus can be
expressed as
\begin{align}\label{eqn: BS receive signal}
\!\!\!\!\!\yb^{\rm U}_0(t) = \sum_{\ell=1}^{L} \gb_{\ell} \sqrt{p^{\rm
        U}_\ell}s_\ell^{\rm U} (t)+ \Hb_{0}(\xb(t)+\ub_{\rm tx}(t)) +\zb^{\rm U}(t),
\end{align}
where the second term in the right hand side (RHS) is the SI and the third term $\zb^{\rm U}(t)$ is the AWGN
following  $\mathcal{CN}(\zerob,\sigma_z^2\Ib_{N_t})$.
The SI power is in general much stronger than the signals transmitted from the UMUs and the AWGN, and
therefore the SI has to be suppressed in order to decode the desired information data properly.
However, simply mitigating the SI in the digital domain is insufficient.
In fact, the SI power could be so large such that the receiver RF chain gets saturated as the dynamics of $\yb^{\rm U}_0(t)$ may be out of the range that the ADC can support \cite{Katti_ACM13}.
In view of this,  SIC has to be carried out in the analog domain before ADC \cite{Katti_ACM13}, as shown in Figure \ref{fig: block diagram}.
Let $ \Hb_{0}= \hat \Hb_{0} +  \Phib_{0}$, where $\hat \Hb_{0}$ and $\Phib_{0}$ respectively represent the channel estimate of $ \Hb_{0}$ and the associated estimation error matrix. Suppose that the analog SIC subtracts the estimated SI signal $\hat \Hb_{0}(\xb(t)+\ub_{\rm tx}(t))$
from $\yb^{\rm U}_0(t)$. Then the signal before ADC is given by
\begin{align}\label{eqn: BS receive signal2}
\!\!\!\!\yb^{\rm U}_1(t) = \sum_{\ell=1}^{L} \gb_{\ell} \sqrt{p^{\rm
        U}_\ell}s_\ell^{\rm U} (t)+ \Phib_{0}(\xb(t)+\ub_{\rm tx}(t)) +\zb^{\rm U}(t).
\end{align}

After ADC, we obtain the following discrete-time signal
\begin{align}\label{eqn: BS receive signal3}
\yb^{\rm U}_2[n] = & \sum_{\ell=1}^{L} \gb_{\ell} \sqrt{p^{\rm
        U}_\ell}s_\ell^{\rm U} [n]+ \Phib_{0}(\xb[n]+\ub_{\rm tx}[n]) \notag \\
    &+ \ub_{\rm rx}[n]+\zb^{\rm U}[n].
\end{align}
Here, $\ub_{\rm rx}[n] $ is the noise caused by the non-ideal receiver RF chain and follows
\begin{align}
\ub_{\rm rx}[n]\sim \mathcal{CN}(\zerob, \beta_2\diag(\E[\yb^{\rm U}_1[n] (\yb^{\rm U}_1[n] )^H]) ),\label{eqn: urx}
\end{align} where $\beta_2 \ll 1$ \cite{Day2012bidirectional}.
Before data detection, digital SIC is further carried out for $\yb^{\rm U}_2[n] $. According to \cite{Katti_14,Katti_ACM13}, it is possible to suppress the linear SI components $ \Phib_{0}\xb[n] $ and non-linear components $ \Phib_{0}\ub_{\rm tx}[n]+  \ub_{\rm rx}[n]$ separately.
For simplicity, we model that the linear SI power  can be reduced by a factor of $\delta_1\ll 1$ and the non-linear SI power can be reduced by a factor of $\delta_2\ll 1$, due to the digital SIC.
Therefore, the signal at the output of the digital SIC is given by
\begin{align}\label{eqn: BS receive signal4}
\yb^{\rm U}_3[n] =& \sum_{\ell=1}^{L} \gb_{\ell} \sqrt{p^{\rm
        U}_\ell}s_\ell^{\rm U} [n]+  \sqrt{\delta_1}\Phib_{0}\xb[n] \notag\\ 
    &+ \sqrt{\delta_2}(\Phib_{0}\ub_{\rm tx}[n]+  \ub_{\rm rx}[n])+\zb^{\rm U}[n].
\end{align}

To detect $s_{\ell}^{\rm U}[n]$, the BS applies a linear receive beamformer, denoted by $\vb_\ell\in
\mathbb{C}^{{MN_R }}$, to $\yb_3^{\rm U}[n]$, for all $\ell\in \Lc$.
The SINR at the output of beamformer for UMU $\ell$ is thus given by
\begin{align}\label{eqn: UMU sinr0}
&{\rm SINR}_\ell^{\rm U}(\Wb,\pb^{\rm U},\vb_\ell) =\notag \\
&\frac{p_\ell^{\rm U}| \vb_\ell^H\gb_\ell|^2 }
{\sum_{j \neq \ell}^L p_j^{\rm U}| \vb_\ell^H\gb_j|^2 +
    \vb_\ell^H \E[ \SigmaB(\Wb,\Phib_{0}) ]\vb_\ell
      + \sigma_z^2\|\vb_\ell\|_2^2},\end{align}
where 
\begin{align}\label{eqn: Sigma W}
    \vb_\ell^H \E[  \SigmaB(\Wb,\Phib_{0}) ]\vb_\ell \triangleq & ~
\delta_1    \vb_\ell^H\E[\Phib_{0}\xb(t)\xb^H(t)\Phib_0^H] \vb_\ell
 \notag \\
&+ \delta_2  \vb_\ell^H\E[\Phib_{0}\ub_{\rm tx}[n]\ub_{\rm tx}^H[n]\Phib_0^H] \vb_\ell  +   \delta_2  \vb_\ell^H \E[\ub_{\rm rx}[n]\ub_{\rm rx}^H[n]]\vb_\ell .
\end{align}
It is shown in Appendix \ref{appx: derivations} that \eqref{eqn: Sigma W} can be compactly expressed as

\vspace{-0.3cm}
{\small\begin{align}
     \vb_\ell^H \E[  \SigmaB(\Wb,\Phib_{0}) ]\vb_\ell&=(\delta_2  \beta_2)  \vb_\ell ^H \bigg( \sum_{j=1}^{L}  p^{\rm
     	U}_j \diag( |\gb_{j}|^2) \bigg)\vb_\ell +  \vb_\ell ^H \Omegab(\Wb, \Rb_{\Phi_0}) \vb_\ell,   \label{eqn: omega0}\\
    &= (\delta_2  \beta_2)  \vb_\ell ^H \bigg( \sum_{j=1}^{L}  p^{\rm
     	U}_j \diag( |\gb_{j}|^2) \bigg)\vb_\ell +  \sum_{k=1}^K \wb_k ^H \LambdaB(\vb_\ell,\Rb_{\Phi_0}) \wb_k,    \label{eqn: lambda0}
\end{align}}
\!\!using the matrices $\Omegab(\Wb, \Rb_{\Phi_0}) $ and $\LambdaB(\vb_\ell,\Rb_{\Phi_0}) $ defined in \eqref{eqn: Omega} and \eqref{eqn: LambdaB}, where
$\Rb_{\Phi_0}\triangleq \E[{\rm vec}(\Phib_0){\rm vec}^H(\Phib_0)]$  is the correlation matrix of vectorized $\Phib_0$.
Let us define
\begin{align}
& \tilde \Gb_j \triangleq \gb_j\gb_j^H +  \delta_2\beta_2 \diag(|\gb_j|^2),\\
& \tilde \sigma_z^2 \triangleq (1+ \delta_2\beta_2 )\sigma_z^2,
\end{align}
and let $\gamma^{\rm U}_\ell>0$ be the SINR target of UMU $\ell$ for all $\ell\in \Lc$.
Then either using \eqref{eqn: omega0} or \eqref{eqn: lambda0}, the SINR constraint ${\rm SINR}_\ell^{\rm U} \geq \gamma^{\rm U}_\ell$ for UMU $\ell$ can be written as
 
\vspace{-0.4cm}
{\small \begin{align}
&{\rm SINR}_{\ell}^{\rm U}(\Wb,\pb^{\rm U},\vb_\ell) \geq \gamma^{\rm U}_\ell \notag \\
&\Leftrightarrow 
\frac{p_\ell^{\rm U}| \vb_\ell^H\gb_\ell|^2 /\rho_\ell^{\rm U}}
{\sum_{j =1 }^L p_j^{\rm U} \vb_\ell^H \tilde \Gb_j \vb_\ell +
	\vb_\ell^H \Omegab(\Wb,\Rb_{\Phib_{0}})\vb_\ell
	+ \tilde \sigma_z^2\|\vb_\ell\|_2^2}  \geq 1, \label{eqn: uplink sinr v} \\
&\Leftrightarrow 
\frac{p_\ell^{\rm U}| \vb_\ell^H\gb_\ell|^2 /\rho_\ell^{\rm U}}
{\sum_{j =1 }^L p_j^{\rm U} \vb_\ell^H \tilde \Gb_j \vb_\ell +
	\sum_{k=1}^K \wb_k^H \LambdaB(\vb_\ell ,\Rb_{\Phib_{0}})\wb_k
	+ \tilde \sigma_z^2\|\vb_\ell\|_2^2}  \geq 1, \label{eqn: uplink sinr w}
\end{align}} \vspace{-0.0cm}
\!\!where $\frac{1}{\rho_\ell^{\rm U} }\triangleq  1+1/\gamma_\ell^{\rm U}$.
It will be seen shortly that both formulations in \eqref{eqn: uplink sinr v} and \eqref{eqn: uplink sinr w} are useful in the development of the proposed algorithms.

\subsubsection{Constraint on ADC Input Signal Power}\label{subsec: ADC constraint}

As mentioned,  the signal dynamics should not be out of the range  that  the ADC can support as, otherwise, signal distortion can occur. 
Therefore, it is required to constrain the signal power of $\yb_1^{\rm U}(t)$ at the input of the ADC. Let $\gamma^{\rm ADC}>0$ denote the maximum tolerable ADC input signal power.
The ADC power constraint for the receiver RF chains are given by 
\begin{align}\label{eqn: max ADC constraint}
 \!\!\!\!\!\!   {\rm ADC} (\Wb,\pb^{\rm U}) \triangleq \diag(\E\{\yb_1^{\rm U}(t)( \yb_1^{\rm U}(t) )^H\})\!\! \preceq\!  \gamma^{\rm ADC} \Ib_{N_t}.
\end{align}
It can be shown that, for $n=1,\ldots,N_t,$
\begin{align}
  &[ {\rm ADC} (\Wb,\pb^{\rm U}) ]_{n,n} =  \eb_n^T\E\{  \Phib_0 \Wb \Wb^H  \Phib_0^H\}  \eb_n \notag\\
  &
  +\beta_1  \eb_n^T \E\{ \Phib_0  \diag(\Wb\Wb^H)  \Phib_0^H \} \eb_n + \sum_{j=1}^L p_j^{\rm U} |\eb_n^T\gb_j|^2  + \sigma_z^2,\notag \\
 &=\sum_{k=1}^K \wb_k^H  \Upsilonb_n(\{ \Rb_{\Phi_0,m}\})  \wb_k + \sum_{j=1}^L p_j^{\rm U} |\eb_n^T\gb_j|^2  + \sigma_z^2, \label{eqn: ADC explict}
\end{align}
where $\Upsilonb_n(\{ \Rb_{\Phi_0,m}\}) \triangleq \bar \Rb_{\Phi_0,n}  + \beta_1  \diag(\{\eb_n^T \Rb_{\Phi_0,m} \eb_n\}_m)$, 
and $\bar \Rb_{\Phi_0,n}$ and $\Rb_{\Phi_0,m}$ are defined in \eqref{eqn: Rn}.


\subsection{Proposed Problem Formulation}\label{subsec: problem formulation}
Our goal is to design the UL transmission powers $\{p_\ell^{\rm U}\}$ and UL and DL beamformers $\{\vb_\ell\}$, $\{\wb_k\}$
so that the transmission power of the network (including the BS and the UMUs) is minimized subject to user SINR constraints and the ADC input power constraint. Mathematically, the QoS-based linear transceiver design problem is formulated as follows
\begin{subequations}\label{eqn: main problem}
\begin{align}
{\sf (P)}~~~    \min_{ \substack{\{\wb_k\},\{ \vb_\ell\}, \\ \{p_\ell^{\rm U}\geq 0\}}}~&\sum_{k=1}^{K}
    \|\wb_k\|_2^2 + \sum_{\ell=1}^L p_\ell^{\rm U} \\
    {\rm s.t.}~~&
     {\rm SINR}_{i}^{\rm D}(\Wb,\pb^{\rm U}) \geq \gamma^{\rm D}_i,~i\in \Kc, \\
     &   {\rm SINR}_{\ell}^{\rm U}(\Wb,\pb^{\rm U},\vb_\ell) \geq \gamma^{\rm U}_\ell,~\ell\in \Lc,\label{eqn: main problem uplink}  \\
     &{\rm ADC} (\Wb,\pb^{\rm U})  \preceq  \gamma^{\rm ADC}\Ib_{N_t},  \label{eqn: main problem ADC}\\
     & \|\vb_\ell\|_2=1,~\ell\in \Lc.
    \end{align}
\end{subequations}
Unfortunately, problem {\sf (P)} is non-convex and difficult to solve. Specifically, the UL SINR constraints and DL SINR constraints are coupled with each other due to the FD BS, which makes {\sf (P)}  drastically different from the traditional design problems in HD systems \cite{Gershman2010_SPM}.
In subsequent sections, we present two methods to handle {\sf (P)}. Firstly, we show that for a special case of {\sf (P)}, the problem is globally solvable in a polynomial-time complexity. Secondly, for general {\sf (P)}, we propose an efficient AO method to solve it approximately.

\subsection{Review of Half-Duplex BF Solutions}\label{subsec: review of HD}
Before studying the methods for solving the FD design problem {\sf (P)}, let us review some existing results about the QoS-based design problems in a HD system. These results will be used in the development of the proposed methods for solving {\sf (P)}.

In the HD system, the BS serves the DMUs and UMUs separately, either in different time slots or over distinct frequency bands. Thus, there is no SI (i.e., the term $\sum_{k=1}^K \wb_k^H \LambdaB(\vb_\ell ,\Phib_{0})\wb_k$ in \eqref{eqn: uplink sinr w}) and no UL-to-DL CCI (i.e., the term $\sum_{j=1}^{L} p_j^{\rm U}|f_{ji}|^2$ in \eqref{eqn: downlink sinr2}).  The ADC input power constraint is also not necessary as the signal power at the input of ADC is generally small in the absence of SI.
Therefore, the DL design problem  {\sf (HD-DL)}
\begin{subequations}\label{eqn: power min problem HD dl}
    \begin{align}
     \min_{ \{\wb_k\} }~&\sum_{k=1}^{K}
    \|\wb_k\|_2^2 \\
    {\rm s.t.}~~&
    \frac{|\hb_{i}^H\wb_{i}|^2/\rho_i^{\rm D}}
    { \sum_{k=1}^K \wb_k^H \tilde \Hb_i \wb_k +
        \sigma_i^2}\geq 1,~k\in \Kc, \label{eqn: power min problem HDD Dl C1}
    \end{align}
\end{subequations}
and UL design problem {\sf (HD-UL)}
\begin{subequations}\label{eqn: power min problem HD ul}
    \begin{align}
      \min_{ \{ \vb_\ell\},  \{p_\ell^{\rm U}\geq 0\} }~&
     \sum_{\ell=1}^L p_\ell^{\rm U} \\
    {\rm s.t.}~~&
   \frac{p_\ell^{\rm U}| \vb_\ell^H\gb_\ell|^2 /\rho_\ell^{\rm U}}
   {\sum_{j =1 }^L p_j^{\rm U} \vb_\ell^H \gb_j\gb_j^H \vb_\ell
    +  \sigma_z^2\|\vb_\ell\|_2^2}  \geq 1,  \label{eqn: power min problem HDD ul C1} \\
    & \|\vb_\ell\|_2=1,~\ell\in \Lc,
    \end{align}
\end{subequations}
can respectively be deduced from {\sf (P)}, and optimized independently \cite{Gershman2010_SPM}.

While both {\sf (HD-DL)} and {\sf (HD-UL)} appear non-convex problems, it is well known that both of them own certain hidden convexity and can be globally solved in a polynomial-time complexity.
Specifically,  {\sf (HD-UL)} is shown equivalent to a convex semidefinite program (SDP). 
\begin{Lemma}\label{lemma: HDU SDP}
	{\rm \cite{LiuHongDaiSPL13}} Suppose that {\sf (HD-UL)} is feasible. Then  {\sf (HD-UL)} is equivalent to the following SDP
		\begin{align}  \label{eqn: HDU SDP}
		\max_{  \{p_\ell^{\rm U}\geq 0\} }~&
		\sum_{\ell=1}^L p_\ell^{\rm U} \\
		{\rm s.t.}~~&
		\sum_{j =1 }^L p_j^{\rm U} \gb_j\gb_j^H
		+ \sigma_z^2\Ib_{N_t} \succeq \bigg(\frac{p_\ell^{\rm U}}{\rho_\ell^{\rm U}}\bigg)\gb_\ell\gb_\ell^H,~\forall \ell\in \Lc,\notag
		\end{align}
\end{Lemma}
and therefore {\sf (HD-UL)} is polynomial-time solvable.
\begin{Lemma}\label{lemma: HDU sol}
	{\rm \cite[Proposition 3.1]{LiuHongDaiSPL13}\cite{SongTCOM07}} Suppose that {\sf (HD-UL)} is feasible. The optimal power solution of problem  {\sf (HD-UL)} is unique and satisfies
	\begin{align}\label{eqn: classical uplink fixed point eqs}
	\bigg(\frac{p_\ell^{\rm U}}{\rho_\ell^{\rm U}}\bigg) \gb_\ell^H\bigg( \sum_{j =1 }^L p_j^{\rm U} 
	 \gb_j\gb_j^H
	+ \sigma_z^2\Ib_{N_t} \bigg)^{-1}\gb_\ell =1,~\forall \ell \in \Lc.
	\end{align}
	The optimal beamformers are given by
	\begin{align}
\!\!\!\!	\vb_\ell =\frac{\vb_\ell}{\|\vb_\ell \|},\tilde \vb_\ell=\bigg( \sum_{j =1 }^L p_j^{\rm U} \gb_j\gb_j^H
	+  \sigma_z^2\Ib_{N_t} \bigg)^{-1}\gb_\ell,~\forall \ell \in \Lc.
	\end{align}
\end{Lemma}
In fact, it can be verified that at the optimum, constraint \eqref{eqn: power min problem HDD ul C1}  must hold with equality, which leads to  \eqref{eqn: classical uplink fixed point eqs}.
Thus, Lemma \ref{lemma: HDU sol} says that the $L$ nonlinear system of equations of  \eqref{eqn: classical uplink fixed point eqs}  has a unique solution, uniquely determined by $\gb_j $'s, $\gb_\ell$'s, $\gamma_\ell^{\rm U}$'s and $\sigma_z^2$. Moreover, the optimal $\{p_\ell^{\rm U}\}$ can be obtained by solving equations \eqref{eqn: classical uplink fixed point eqs}  using a fixed-point method \cite{Liu1998,Wiesel2006}.

The DL problem {\sf (HD-DL)} can be solved by considering an equivalent second-order cone program (SOCP) \cite{Wiesel2006}
or a SDP \cite{BK:Bengtsson01} which is obtained by a semidefinite relaxation (SDR) technique \cite{Luo2010_SPM}.
Both SOCP and SDP are convex problems and are efficiently solvable by off-the-shelf solvers.
The fixed-point method for UL problems can also be used to solve the DL problem {\sf (HD-DL)}, through a powerful UDD \cite{Boche2002,Wiesel2006,Gershman2010_SPM}.
\begin{Lemma}\label{lemma: HDD sol}
    {\rm \cite{Gershman2010_SPM}}  Suppose that {\sf (HD-DL)} is feasible.  
    {\sf (HD-DL)} has a virtual UL counterpart as follows
    \begin{subequations}\label{eqn: HDD virtual uplink}
        \begin{align}
        \min_{ \{ \tilde \wb_i\},  \{\lambda_i\geq 0\} }~&
        \sum_{i=1}^K \lambda_i \sigma_i^2 \\
        {\rm s.t.}~~&
        \frac{\lambda_i| \tilde \wb_i^H\hb_i|^2 /\rho_i^{\rm D}}
        {\sum_{k =1 }^K \lambda_k \tilde \wb_i^H \tilde \Hb_k \tilde \wb_i
            +  \|\tilde \wb_i\|_2^2}  \geq 1,  \label{eqn: power min problem HD ul C1} \\
        & \|\tilde \wb_i\|_2=1,~i \in \Kc,
        \end{align}
    \end{subequations}
which has the same optimal objective value as {\sf (HD-DL)}.
\end{Lemma}
In Section \ref{subsec: UD duality}, we will show that such UDD can be generalized to the FD system.


\section{Optimal Solution for {\sf (P)} with i.i.d. SI Channel Estimation Error }\label{sec: bisection}

In this section, we consider a special case of {\sf (P)} by assuming that elements of the SI channel estimation error are i.i.d. Specifically, $\Rb_{\Phi_0}$ is assumed to be $\Rb_{\Phi_0}=\sigma_{\Phi_0}^2 \Ib_{N_t^2}$ for some $\sigma_{\Phi_0}^2>0$.
Under this assumption, one can show that $\Omegab(\Wb, \Rb_{\Phib_0}) $ in the denominator of \eqref{eqn: uplink sinr v} reduces to $\xi \sum_{i=1}^K\|\wb_i\|_2^2 \Ib_{N_t}$, where $\xi\triangleq (\delta_1+\delta_2\beta_1 + \delta_2\beta_2(1+\beta_1)) \sigma_{\Phi_0}^2$, and $\Upsilonb_n(\{ \Rb_{\Phi_0,m}\})$ in the ADC input power constraint \eqref{eqn: ADC explict}
reduces to $\Upsilonb_n(\{ \Rb_{\Phi_0,m}\}) =
\sigma_{\Phi_0}^2 (1+\beta_1)\sum_{i=1}^K\|\wb_i\|_2^2 \Ib_{N_t}$. As a result,
problem {\sf (P)} simplifies to
\begin{subequations}\label{eqn: power min problem}
	\begin{align}
	{\sf (P1)}~ &  \min_{ \substack{\{\wb_k\},\{ \vb_\ell\}, \\ \{p_\ell^{\rm U}\geq 0\}}}~\sum_{k=1}^{K}
	\|\wb_k\|_2^2 + \sum_{\ell=1}^L p_\ell^{\rm U} \\
	{\rm s.t.}~~&
	\frac{|\hb_{i}^H\wb_{i}|^2/\rho_i^{\rm D}}
	{ \sum_{k=1}^K \wb_k^H \tilde \Hb_i \wb_k +
			\hat \sigma_i^2(\pb^{\rm U})}\geq 1,~i\in \Kc, \\
	&   \frac{p_\ell^{\rm U}| \vb_\ell^H\gb_\ell|^2 /\rho_\ell^{\rm U}}
	{\sum_{j =1 }^L p_j^{\rm U} \vb_\ell^H \tilde \Gb_j \vb_\ell +
		\xi\sum_{k=1}^K \|\wb_k\|_2^2
		+ \tilde \sigma_z^2 } \geq 1,~\ell\in \Lc,  \label{eqn: power min problem C2}\\
	&
	\overline  {\rm ADC} (\Wb,\pb^{\rm U})  \preceq  \gamma^{\rm ADC}\Ib_{N_t},\\
	& \|\vb_\ell\|_2=1,~\ell\in \Lc,
	\end{align}
\end{subequations}
where $[\overline {\rm ADC} (\Wb,\pb^{\rm U}) ]_{n,n} \triangleq \sigma_{\Phi_0}^2 (1+\beta_1)\sum_{i=1}^K\|\wb_i\|_2^2 + \sum_{j=1}^L p_j^{\rm U} |\eb_n^T\gb_j|^2  + \sigma_z^2,$ for all $n=1,\ldots,N_t.$
Note that even when $\Rb_{\Phi_0} \neq \sigma_{\Phi_0}^2 \Ib_{N_t^2}$, one can still obtain a similar formulation as {\sf (P1)} by considering a worst-case SI. In particular, one can show that 
$\Omegab(\Wb, \Rb_{\Phib_0}) $ is upper bounded as $\Omegab(\Wb, \Rb_{\Phib_0})  \preceq
(\delta_1+\delta_2\beta_1 + \delta_2\beta_2(1+\beta_1))\lambda_{\max}(\Rb_{\Phib_0}) \sum_{k=1}^K \|\wb_k\|_2^2~\Ib_{N_t}$. Besides, $\Upsilonb_n(\{ \Rb_{\Phi_0,m}\})$ is upper bounded as
$\Upsilonb_n(\{ \Rb_{\Phi_0,m}\}) \preceq
\lambda_{\max}(\Rb_{\Phi_0})(1+\beta_1)\sum_{i=1}^K\|\wb_i\|_2^2 \Ib_{N_t}$.
Therefore, by replacing $\Omegab(\Wb, \Rb_{\Phib_0}) $ and $\Upsilonb_n(\{ \Rb_{\Phi_0,m}\})$ by their respective upper bounds, one can arrive at a similar formulation as {\sf (P1)}.
Interestingly, as we will show shortly, despite that {\sf (P1)} is still a non-convex problem, it is actually polynomial-time solvable.

\subsection{A Globally Solvable Non-Convex Subproblem}
One of the keys to the developed algorithm for solving {\sf (P1)} is 　to consider the following problem
\begin{subequations}\label{eqn: power min problem eta}
    \begin{align}
    {\sf (P_\eta)}~~~~&F(\eta) =  \min_{ \substack{\{\wb_k\},\{ \vb_\ell\}, \\ \{p_\ell^{\rm U}\geq 0\}}}~\sum_{k=1}^{K}
    \|\wb_k\|_2^2 + \sum_{\ell=1}^L p_\ell^{\rm U} \\
    {\rm s.t.}~~&
    \frac{|\hb_{i}^H\wb_{i}|^2/\rho_i^{\rm D}}
    { \sum_{k=1}^K \wb_k^H \tilde \Hb_i \wb_k +
        	\hat \sigma_i^2(\pb^{\rm U}) }\geq 1,~i\in \Kc,   \label{eqn: power min problem eta C1} \\
    &   \frac{p_\ell^{\rm U}| \vb_\ell^H\gb_\ell|^2 /\rho_\ell^{\rm U}}
    {\sum_{j =1 }^L p_j^{\rm U} \vb_\ell^H \tilde \Gb_j \vb_\ell +
        \xi\eta
        + \tilde \sigma_z^2 } \geq 1,~\ell\in \Lc,   \label{eqn: power min problem eta C2}  \\
    &\overline {\rm ADC} (\Wb,\pb^{\rm U})  \preceq  \gamma^{\rm ADC}\Ib_{N_t},  \label{eqn: power min problem eta C3} \\
    & \|\vb_\ell\|_2=1,~\ell\in \Lc,
    \end{align}
\end{subequations}
where $F(\eta) $ denotes the optimal objective value.
The difference between  {\sf (P$_\eta$)} and  {\sf (P1)}  is that the term $\sum_{k=1}^K \|\wb_k\|_2^2$ in \eqref{eqn: power min problem C2} is replaced by a parameter $\eta>0$.
Intriguingly, problem {\sf (P$_{\eta}$)}, though not being convex, is polynomial-time solvable.
\begin{Proposition}\label{prop: two stage}
    Suppose that problem {\sf (P$_\eta$)}  is feasible. Then, the optimal $\{\vb_\ell,p_\ell^{\rm U}\}$ of {\sf (P$_\eta$)}, denoted by $\{\vb_\ell(\eta),p_\ell^{\rm U}(\eta)\}$, is the solution to the following UL problem
    \begin{subequations}\label{eqn: F uplink}
        \begin{align}
        &\{\vb_\ell(\eta),p_\ell^{\rm U}(\eta)\}_{\ell=1}^L =\arg   \min_{ \substack{\{ \vb_\ell\},  \{p_\ell^{\rm U}\geq 0\}}}~ \sum_{\ell=1}^L p_\ell^{\rm U} \\
        &{\rm s.t.}~~
           \frac{p_\ell^{\rm U}| \vb_\ell^H\gb_\ell|^2 /\rho_\ell^{\rm U}}
        {\sum_{j =1 }^L p_j^{\rm U}  \vb_\ell^H \tilde \Gb_j \vb_\ell +
            (\xi\eta + \tilde \sigma_z^2)}  \geq 1,~\ell\in \Lc, \\
        &~~~~~ \|\vb_\ell\|_2=1,~\ell\in \Lc.
        \end{align}
    \end{subequations}
    and the optimal $\{\wb_k\}$ of {\sf (P$_\eta$)}, denoted by $\{\wb_k(\eta)\}$, is the solution to the following DL problem
    \begin{subequations}\label{eqn: F downlink}
        \begin{align}
       & \{\wb_k(\eta)\}_{k=1}^K=  \arg  \min_{ \substack{\{\wb_k\} }}~\sum_{k=1}^{K}
        \|\wb_k\|_2^2  \\
       & {\rm s.t.}~~
        \frac{|\hb_{i}^H\wb_{i}|^2/\rho_i^{\rm D}}
        { \sum_{k=1}^K \wb_k^H \tilde \Hb_i \wb_k +
           	\hat \sigma_i^2(\pb^{\rm U}(\eta)) }\geq 1,~i\in \Kc,   \label{eqn: F downlink C1} \\
        &~~~~ ~\overline {\rm ADC} (\Wb,\pb^{\rm U}(\eta))  \preceq  \gamma^{\rm ADC}\Ib_{N_t}. \label{eqn: F downlink C2}
        \end{align}
    \end{subequations}
    Both problems in \eqref{eqn: F uplink} and \eqref{eqn: F downlink} are polynomial-time solvable.
\end{Proposition}

{\bf Proof:} 
Note that the UL SINR constraints \eqref{eqn: power min problem eta C2} must hold with equality at the optimum; otherwise, one can further reduce $p_\ell^{\rm U}$'s without violating the other constraints. Since $\{\vb_\ell\}$ appears only in  \eqref{eqn: power min problem eta C2},   $\{\vb_\ell(\eta)\}$ is the maximum SINR solution given by
\begin{align}
\vb_\ell(\eta)&=\frac{\vb_\ell(\eta)}{\|\vb_\ell (\eta)\|_2}, \\
\tilde \vb_\ell(\eta)&=\bigg( \sum_{j =1 }^L p_j^{\rm U}(\eta) \tilde \Gb_j
+(\xi\eta+ \tilde \sigma_z^2)\Ib_{N_t} \bigg)^{-1}\gb_\ell,~\forall \ell \in \Lc. \notag
\end{align}
So, $\{p_\ell^{\rm U}(\eta)\}$ satisfies, $\forall \ell \in \Lc,$
\begin{align}\label{eqn: uplink sol}
\!\!\!\!\! \frac{p_\ell^{\rm U}(\eta)}{\rho_\ell^{\rm U}} \gb_\ell^H\bigg( \sum_{j =1 }^L p_j^{\rm U}(\eta) \tilde \Gb_j
+(\xi\eta + \tilde \sigma_z^2)\Ib_{N_t} \bigg)^{-1}\gb_\ell =1.
\end{align}
By applying Lemma \ref{lemma: HDU sol}, we see that $\{p_\ell^{\rm U}(\eta)\}$ is uniquely determined by \eqref{eqn: uplink sol}, regardless of the constraints in  \eqref{eqn: power min problem eta C1} and  \eqref{eqn: power min problem eta C3}. As a result, $\{\vb_\ell(\eta),p_\ell^{\rm U}(\eta)\}$ essentially can be obtained by solving \eqref{eqn: F uplink}, which is similar to  the HD UL problem {\sf (HD-UL)} and is polynomial-time solvable.
Once $\{\vb_\ell(\eta),p_\ell^{\rm U}(\eta)\}$  are given, it is clear that $\{\wb_k(\eta)\}$ can be obtained by solving \eqref{eqn: F downlink}.
Problem \eqref{eqn: F downlink} is polynomial-time solvable as, similar to {\sf (HD-DL)}, \eqref{eqn: F downlink}  can be formulated as an SOCP \cite{Wiesel2006} or  solved by SDR \cite{BK:Bengtsson01}.
\hfill $\blacksquare$

Proposition \ref{prop: two stage} is constructive as it provides a two-stage approach to solving {\sf (P$_\eta$)}.
In the next subsection, we analyze the relation between {\sf (P$_\eta$)} and {\sf (P1)}. Based on these results, we develop a bisection strategy to solve {\sf (P1)} to global optimality. 

\subsection{Proposed Bisection Algorithm for Solving {\sf (P1)}}\label{subsec: bisection}

\begin{figure}[!t]
	\begin{center}
		{\resizebox{.3\textwidth}{!}
			{\includegraphics{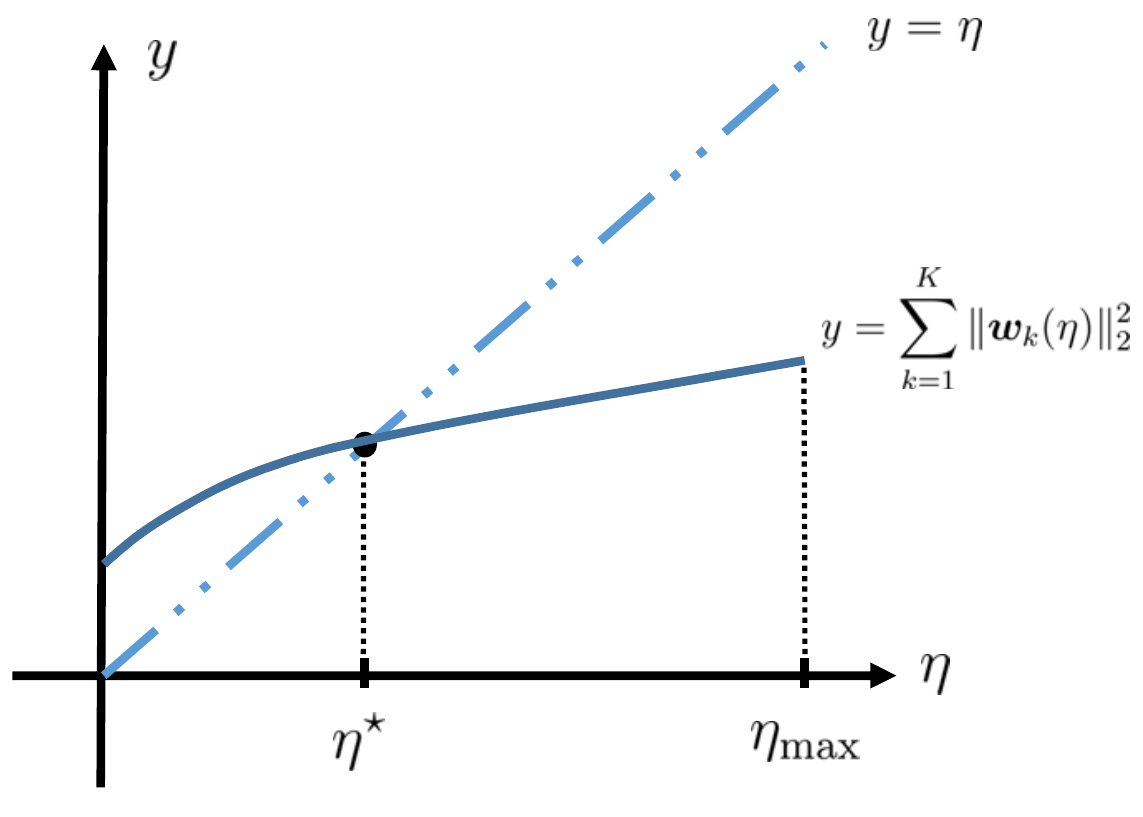}}}
	\end{center}\vspace{-0.5cm}
	\caption{Illustration of  Proposition \ref{prop: eta nd P_eta 2}. It is assumed that {\sf (P1)} is feasible.}
	\vspace{-0.3cm}\label{fig: y}
\end{figure}
 
We first have the following proposition.

\begin{Proposition}\label{property: eta and P_eta}
    Suppose that {\sf (P1)} is feasible. Let $\{\wb_k^\star\}$ be the optimal DL beamformers of   {\sf (P1)}  and let $\eta^\star \triangleq \sum_{k=1}^K \|\wb_k^\star\|_2^2 $.
  Then\\
    (a) $\eta^\star$ is unique; \\
    (b) When $\eta=\eta^\star$, $\{\wb_k(\eta)\}$ of {\sf (P$_\eta$)} is optimal to {\sf (P1)}
    and satisfies $\sum_{k=1}^K \|\wb_k(\eta)\|_2^2=\eta^\star$; \\
    (c) 
    $F(\eta)$ of {\sf (P$_\eta$)} is monotonically increasing with $\eta$.
\end{Proposition}

{\bf Proof:}  See Appendix \ref{appx: proof of prop 2}. \hfill $\blacksquare$

Proposition \ref{property: eta and P_eta} suggests that one can solve {\sf (P1)} by searching the unique optimal $\eta^\star$ of  {\sf (P$_\eta$)}.
The following result provides important insight into how $\eta$  can be searched efficiently.

\begin{Proposition}\label{prop: eta nd P_eta 2}
    Suppose that {\sf (P1)} is feasible and that {\sf (P$_\eta$)} is feasible for $\eta\in [0,\eta_{\max}]$, where $\eta_{\max}>0$. Then  \\
    (a) $\sum_{k=1}^K \|\wb_k(\eta)\|_2^2$  of {\sf (P$_\eta$)} is concave and increasing with respect to $\eta\in [0,\eta_{\max}]$. \\
    (b) $\sum_{k=1}^K \|\wb_k(\eta)\|_2^2 > \eta$ if  and only if $\eta < \eta^\star$. \\
   Moreover, {\sf (P1)} is infeasible if and only if {\sf (P$_\eta$)} has $\sum_{k=1}^K \|\wb_k(\eta)\|_2^2 > \eta$ for all $\eta\in [0,\eta_{\max}]$.
\end{Proposition}

{\bf Proof:}  See Appendix \ref{appx: proof of prop 3}. \hfill $\blacksquare$

From Proposition \ref{prop: eta nd P_eta 2}, we can visualize the function $\sum_{k=1}^K \|\wb_k(\eta)\|_2^2$ as in Figure \ref{fig: y}, provided that {\sf (P1)} is feasible.
Therefore, one can search the optimal $\eta^\star$ in a bisection fashion, by comparing the values of $\sum_{k=1}^K \|\wb_k(\eta)\|_2^2$ 
and $\eta$. Specifically, if $\sum_{k=1}^K \|\wb_k(\eta)\|_2^2 > \eta$, $\eta$ should be increased, whereas if $\sum_{k=1}^K \|\wb_k(\eta)\|_2^2 \leq \eta$, then $\eta$ should be decreased.
Moreover, if it happens that $\sum_{k=1}^K \|\wb_k(\eta)\|_2^2 > \eta$ for all $\eta\in [0, P_{\max}]$, where $P_{\max}$ denotes some bisection upper bound, then 
it will make $\eta$ converges to the bisection upper bound and thus one can declare that {\sf (P1)} is infeasible for $\sum_{k=1}^K \|\wb_k\|_2^2 \leq P_{\max}$.
Based on these observations, we develop a bisection method for solving {\sf (P1)} in Algorithm \ref{table: ILS for P1 bisection}.
In Algorithm \ref{table: ILS for P1 bisection}, the parameter $P_{\max}$ denotes the bisection upper bound and is the maximum DL transmission power budget. 

 \begin{algorithm}[h!]
    \caption{Proposed bisection algorithm for solving {\sf (P1)}. }
    \begin{algorithmic}[1]\label{table: ILS for P1 bisection}
        \STATE {\bf Set} $\eta^{\ell b}=0$ and $\eta^{ub}=P_{\max}$.

        \REPEAT
        \STATE Set $\eta^m  \leftarrow \frac{\eta^{ub} + \eta^{\ell b}}{2}.$

        \STATE  Solve {\sf (P$_\eta$)} by solving \eqref{eqn: F uplink} and \eqref{eqn: F downlink} in order.

        \STATE  If {\sf (P$_\eta$)} is infeasible or  if {\sf (P$_\eta$)} is feasible and  $\sum_{k=1}^K \|\wb_k(\eta^m)\|_2^2 \leq \eta^m$, set $\eta^{ub} \leftarrow \eta^m$; otherwise
        set $\eta^{\ell b} \leftarrow \eta^m$.
        

        \UNTIL $|\eta^{ub}-\eta^{\ell b}| \leq \epsilon$, where $\epsilon$ is a preset value.
        If $|\eta^{ub}-P_{\max}|\leq  \epsilon$, then declare that {\sf (P1)} is infeasible.
    \end{algorithmic}
 \end{algorithm}

Since both the bisection algorithm and the subproblems \eqref{eqn: F uplink} and \eqref{eqn: F downlink}  have polynomial-time complexity,
{\sf (P1)} is polynomial-time solvable.

\begin{Theorem}
	Suppose that {\sf (P1)} is feasible. Then Algorithm \ref{table: ILS for P1 bisection} globally solves  {\sf (P1)} in polynomial time.
\end{Theorem}

 From the proof of Proposition \ref{prop: eta nd P_eta 2}(a), it is worth noting that not only $\sum_{k=1}^K \|\wb_k(\eta)\|_2^2$ but also $\sum_{\ell=1}^L p_\ell^{\rm U}(\eta)$ are concave functions of $\eta$.
When applying this insight to the original problem {(\sf P)}, it implies that the objective function $\sum_{k=1}^K \|\wb_k\|_2^2+\sum_{\ell=1}^L p_\ell^{\rm U}$ is concave function of the SI power. Therefore, for such minimization problem, it is unlikely to solve {(\sf P)} jointly with respect to all the variables $\{\wb_k,\vb_\ell, p_\ell^{\rm U}\}$ in a convex manner.
In light of this, we resort to approximation methods in the next section.

\section{KKT Solutions to {\sf (P)} via Alternating
Optimization} \label{sec: AO}

In this section, we consider the general design problem {\sf (P)} in
\eqref{eqn: main problem} and propose a suboptimal method to handle {\sf (P)} by alternating optimization.
In the first subsection, we  present the proposed AO algorithm and show that the algorithm is guaranteed to converge to a Karush-Kuhn-Tucker (KKT) point of {\sf (P)}.
In the second subsection, we generalize the UDD in HD systems \cite{Boche2002,SongTCOM07} to the FD system and use it to develop a computationally efficient  fixed-point based AO algorithm.

\subsection{Proposed AO Algorithm}\label{subsec: AO}

The proposed AO algorithm shares a similar strategy as the iterative algorithms proposed in \cite{VisotskyVTC99,WongZNVTC05,LiuDaiLuoTSP13} for joint transmit and receive beamforming optimization in HD MIMO systems. For the considered problem  {\sf (P)}, we observe that, when the
UL beamformer $\{\vb_\ell\}$ are fixed, {\sf (P)} can be recast as a convex SOCP, through simple change of variables. Moreover, when
the DL beamformer $\{\wb_k\}$ and UL power
$\{p_\ell^{\rm U}\}$ are fixed, $\{\vb_\ell\}$ has a simple
closed-form expression. 

Specifically, let us assume that the UL beamforming vectors $\{\vb_\ell\}$ are fixed in {\sf (P)}. We have
\begin{subequations}\label{eqn: prob formulation AO}
    \begin{align}
    &\min_{ \substack{\{\wb_k\}, \{p_\ell^{\rm U}\geq 0\}}}~\sum_{k=1}^{K}
    \|\wb_k\|_2^2 + \sum_{\ell=1}^L p_\ell^{\rm U} \\
   & {\rm s.t.}~~
    \frac{|\hb_{i}^H\wb_{i}|^2 /{ \rho_i^{\rm D}} }
    { \sum_{k=1}^K \wb_k^H \tilde \Hb_i \wb_k +
              \hat  \sigma_i^2(\pb^{\rm U})}\geq 1,~\forall i\in \Kc, \label{eqn: prob formulation AO_cons1} \\
    &~~   \frac{p_\ell^{\rm U}| \vb_\ell^H\gb_\ell|^2 /\rho_\ell^{\rm U}}
    {\sum_{j =1 }^L p_j^{\rm U} \vb_\ell^H \tilde \Gb_j \vb_\ell +
         \sum_{k=1}^K \wb_k^H \LambdaB(\vb_\ell ,\Rb_{\Phib_{0}})\wb_k
        + \tilde \sigma_z^2 } \geq 1,~\forall \ell\in \Lc,  \label{eqn: prob formulation AO_cons2}\\
       &~~{\rm ADC} (\Wb,\pb^{\rm U})  \preceq  \gamma^{\rm ADC}\Ib_{N_t}.  \label{eqn: prob formulation AO_cons3}
    \end{align}
\end{subequations}
Note from \eqref{eqn: ADC explict} that \eqref{eqn: prob formulation AO_cons3} are convex constraints. Besides, if $\{\wb_k^\star\}$ is an optimal solution to \eqref{eqn: prob formulation AO}, then any phase rotated version of $\{\wb_k^\star\}$ is still an optimal solution. 
Let us consider in \eqref{eqn: prob formulation AO} the change of variables $q_\ell=\sqrt{p_\ell^{\rm U}}$, $\ell\in \Lc$, and apply proper phase rotation to $\{\wb_k\}$ so that $\hb_{k}^H\wb_{k}$ is real-valued for all $k\in \Kc$. Then one can equivalently write \eqref{eqn: prob formulation AO} as follows

\vspace{-0.6cm}
 {\small
 	\begin{subequations}\label{eqn: SOCP}
    \begin{align}
   & \min_{ \substack{\{\wb_k\},\{q_\ell\geq 0\}}}~\sum_{k=1}^{K} \|\wb_k\|^2 + \sum_{\ell=1}^L q_\ell^2 \\
   & {\rm s.t.}~~ \frac{\hb_{i}^H\wb_{i}} {\sqrt{\rho_i^{\rm D}}} \geq \sqrt{\sum_{k=1}^K \| \tilde \Hb_i^{\frac{1}{2}} \wb_k\|_2^2 +
        \sum_{j=1}^{L} q_j^2|f_{ji}|^2 +
        \sigma_i^2},~k\in \Kc,  \\ 
    & \frac{q_\ell |\vb_\ell^H\gb_\ell | }{\sqrt{ \rho^{\rm U}_\ell}} \geq \sqrt{\sum_{j =1 }^L q_j^{2} \vb_\ell^H \tilde
\Gb_j \vb_\ell +
         \sum_{k=1}^K \| \LambdaB^{\frac{1}{2}}(\vb_\ell ,\Rb_{\Phib_{0}})\wb_k\|_2^2
        + \tilde \sigma_z^2 },~\forall \ell\in \Lc, \\
&{\rm ADC} (\Wb,\pb^{\rm U})  \preceq  \gamma^{\rm ADC}\Ib_{N_t}.
    \end{align}
\end{subequations}}
\!\!Problem \eqref{eqn: SOCP} is an SOCP which
can be solved by standard convex solvers.

On the other hand, suppose that $\{\wb_k\}$ and  $\{p_\ell^{\rm U}\}$s are fixed in {\sf (P)}.
Then the optimal $\vb_\ell$ that maximizes ${\rm SINR}_\ell^{\rm U}$ in \eqref{eqn: main problem uplink} (i.e.,  \eqref{eqn: uplink sinr v}) is the maximum SINR beamformer 
\begin{align}\label{eqn: receive msinr}
  \vb_\ell= \frac{\Mb_\ell^{-1} \gb_\ell}{\|\Mb_\ell^{-1} \gb_\ell\|_2}, 
\end{align}
where
$
\Mb_\ell\triangleq \sum_{j =1 }^L p_j^{\rm U} \tilde \Gb_j +
\Omegab(\Wb ,\Rb_{\Phib_{0}})
+ \tilde \sigma_z^2 \Ib_{N_t}. 
$

As a result, we propose to handle {\sf (P)} via updating $(\{\wb_k\}, \{p_\ell^{\rm U}\})$ by solving \eqref{eqn: SOCP} and updating $\{\vb_\ell\}$ by \eqref{eqn: receive msinr} in an alternating fashion, as shown in Algorithm
\ref{table: Algorithm 1}. If given the initial beamforming vectors $\{\vb_\ell^{(0)}\}$, problem \eqref{eqn: SOCP} (i.e., \eqref{eqn: prob formulation AO}) is infeasible, then the algorithm shall declare infeasibility.
However, this does not necessarily imply that {\sf (P)} is infeasible. When  \eqref{eqn: prob formulation AO} is feasible given the initial $\{\vb_\ell^{(0)}\}$, one can show that 
the AO algorithm converges to a KKT solution of  {\sf (P)}.

\begin{algorithm}[t!]
	\caption{Proposed SOCP-based AO algorithm for solving {\sf (P)} in \eqref{eqn: main problem}.}
	\begin{algorithmic}[1]\label{table: Algorithm 1}
		\STATE {\bf Given} initial UL beamforming vectors $\vb_\ell^{(0)}$, $\ell=1,\ldots, L$; set $t\leftarrow 0.$
		Declare infeasibility if \eqref{eqn: SOCP} is infeasible given $\{\vb_\ell^{(0)}\}$.
		
		\REPEAT
		\STATE Given $\{\vb_\ell^{(t)}\}$, obtain $(\{\wb_k^{(t+1)}\}, \{(p_\ell^{\rm U})^{(t+1)}\})$ by solving \eqref{eqn: SOCP} (i.e., the SOCP reformulation of \eqref{eqn: prob formulation AO}).
		\STATE Given $(\{\wb_k^{(t+1)}\}, \{(p_\ell^{\rm U})^{(t+1)}\})$, obtain $\{\vb_\ell^{(t+1)}\}$ by \eqref{eqn: receive msinr}.
		\STATE $t \leftarrow t+1.$
		\UNTIL a predefined convergence condition is satisfied.
	\end{algorithmic}
\end{algorithm}

\begin{Theorem}\label{prop: AO_ud duality}
    Suppose that \eqref{eqn: prob formulation AO} is feasible given initial $\vb_\ell^{(0)}$, $\ell\in
    \Lc$. For Algorithm \ref{table: Algorithm 1}, the total power $\sum_{k=1}^{K} \|\wb_k^{(t)}\|_2^2 + \sum_{\ell=1}^L (p_\ell^{\rm U})^{(t)}$ is non-increasing with the iteration number $t$ and converges as $t\to \infty.$
    Moreover, any limit point of $(\{\wb_k^{(t+1)}\}, \{(p_\ell^{\rm U})^{(t+1)}\}, \{\vb_\ell^{(t+1)}\})$ is a KKT point of  {\sf (P)}.
\end{Theorem}

{\bf Proof:} It is easy to show that the objective value $\sum_{k=1}^{K} \|\wb_k^{(t)}\|_2^2 + \sum_{\ell=1}^L (p_\ell^{\rm U})^{(t)}$ is non-increasing.
The proof of $(\{\wb_k^{(t+1)}\}, \{(p_\ell^{\rm U})^{(t+1)}\}, \{\vb_\ell^{(t+1)}\})$ converging to a KKT point of problem {\sf (P)} can follow \cite[Section IV]{ShiQJ16}.
Firstly, analogous to \cite[Lemma 4]{ShiQJ16} and using \eqref{eqn: SOCP}, one can show that problem \eqref{eqn: prob formulation AO} has
a unique solution up to a phase rotation to $\{\wb_k\}.$ Then, following a similar argument as in  \cite[Proposition 1]{ShiQJ16}, one can prove that any limit point of
 $(\{\wb_k^{(t+1)}\}, \{(p_\ell^{\rm U})^{(t+1)}\}, \{\vb_\ell^{(t+1)}\})$ is a KKT point of {\sf (P)}. The details are omitted here.
\hfill $\blacksquare$


\subsection{FD UDD and Fixed Point Method for Solving \eqref{eqn: prob formulation AO} }\label{subsec: UD duality}

In this subsection, we show that the FD problem \eqref{eqn: prob formulation AO} has a duality property that resembles the UDD (i.e., Lemma \ref{lemma: HDD sol}) in HD systems.
Moreover, similar to the fixed-point method for solving the HD problems  \cite{Liu1998,Wiesel2006}, the FD problem \eqref{eqn: prob formulation AO} can also be solved by efficient fixed-point iterations. To illustrate these methods, let us consider a partial Lagrange dual problem of \eqref{eqn: prob formulation AO}
  	\begin{align}\label{eqn: power min problem dual ADC}
	\max_{\nub\succeq \zerob}~ \psi(\nub)+ (\sigma_z^2 - \gamma^{\rm ADC})\oneb^T\nub
	\end{align}
where $\psi(\nub)$ is given by 
 \begin{subequations}\label{eqn: prob formulation AO dualADC_2}
\begin{align}
{\sf (P2)}~~\psi(\nub)= \min_{ \substack{\{\wb_k\}, \{p_\ell^{\rm U}\geq 0\}}}~&\sum_{k=1}^{K}
\wb_k^H \Bb\wb_k  + \sum_{\ell=1}^L p_\ell^{\rm U} b_\ell 
\\
{\rm s.t.}~~& \eqref{eqn: prob formulation AO_cons1},~\eqref{eqn: prob formulation AO_cons2}.
\end{align}
\end{subequations}
Here, $\nub=[\nu_1,\ldots,\nu_{N_t}]^T\in \mathbb{R}^{N_t}$ contain the dual variables associated with the ADC input power constraint \eqref{eqn: prob formulation AO_cons3};
for notational simplicity, it is also defined that $\Bb\triangleq \Ib_{N_t} + \sum_{n=1}^{N_t} \nu_n\Upsilonb_n(\{ \Rb_{\Phi_0,m}\}) $ and $b_\ell\triangleq 1+ \sum_{n=1}^{N_t} \nu_n |\eb_n^T\gb_j|^2$
for all $\ell \in \Lc.$
Owing to the hidden convexity of problem  \eqref{eqn: prob formulation AO} (i.e., \eqref{eqn: SOCP}), one can show that problem  \eqref{eqn: prob formulation AO} in fact has a zero duality gap (see, e.g., \cite[Proposition 1]{yu2007transmitter}). Therefore, solving the dual problem \eqref{eqn: power min problem dual ADC} is equivalent to solving problem  \eqref{eqn: prob formulation AO}.
As a standard approach, one can solve problem \eqref{eqn: power min problem dual ADC}  by the subgradient method \cite{Boydsubgradient}.
In each update of the subgradient method, one has to solve the subproblem {\sf (P2)}.

The following proposition shows that  {\sf (P2)} has an equivalent duality problem:
\begin{Proposition}\label{prop: FD UDD}
	Suppose that problem {\sf (P2)} in \eqref{eqn: prob formulation AO dualADC_2} is feasible. 
    Then {\sf (P2)} achieves the same optimal objective value as the following problem
	 \begin{subequations}\label{eqn: DMU dual problem}
	 	\begin{align}
	 	&(\{\tilde \wb_k^\star\},\lambdab^\star,\mub^\star) =\arg	\min_{\substack{\{\tilde \wb_k\}, \lambdab, \mub \succeq \zerob}}  ~ 
	 \lambdab^T\sigmab^2+ \tilde \sigma_z^2 \mub^T \oneb   \label{eqn: DMU dual problem target} \\
	 &	{\rm s.t.}~~\frac{\lambda_{k}
	 		|\tilde \wb_k^H\hb_{k}|^2/\rho_{k}^{\rm D} } { \tilde \wb_k^H \Qb(\lambdab,\mub)
	 		\tilde
	 		\wb_k } \geq 1,~k\in \Kc, \label{eqn: DMU dual problem VDL}
 		\end{align}
 		\begin{align}
	 	&~~~~~\frac{\mu_{\ell}
	 		|\vb_\ell^H\gb_\ell|^2/\rho_{\ell}^{\rm U}  } {\sum_{j=1}^L \mu_{j}
	 		\vb_\ell^H \tilde \Gb_j
	 		\vb_\ell+\sum_{k=1}^{K}\lambda_{k} |f_{\ell k}|^2+b_\ell
	 	} \geq  1, \ell\in \Lc,
	 	\label{eqn: DMU dual problem VUL} \\
	 	&~~~~~ \|\tilde \wb_k\|_2=1,~k\in \Kc,
	 	\end{align}
	 \end{subequations}
	 where $\sigmab^2 \triangleq
	 [\sigma_1^2,\ldots,\sigma_K^2]^T$ and
	$ \Qb(\lambdab,\mub)\triangleq  \sum_{i = 1}^{K} \lambda_{i}\tilde
	 \Hb_i+ \sum_{\ell=1}^L \mu_\ell  \LambdaB(\vb_\ell
	 ,\Rb_{\Phib_{0}}) + \Bb.$
	 Moreover, the optimal beamforming direction of {\sf (P2)} can be obtained via \eqref{eqn: DMU dual problem} as
	 \begin{align}\label{eqn: downlink_beam_from_dual dualADC_2}
	 \frac{\wb_k^\star}{\|\wb_k^\star\|_2}=\tilde \wb_k^\star= \frac{\Qb^{-1}(\lambdab^\star,\mub^\star) \hb_k }{\|\Qb^{-1}(\lambdab^\star,\mub^\star) \hb_k\|_2},~k\in \Kc,
	 \end{align}
	 where $\{\wb_k^\star\}$ denotes the optimal $\{\wb_k\}$ of {\sf (P2)}.
\end{Proposition}

{\bf Proof:} See Appendix \ref{appx: proof of FD UDD}. \hfill $\blacksquare$

By comparing \eqref{eqn: DMU dual problem VDL} and \eqref{eqn: DMU
dual problem VUL} with  \eqref{eqn: prob formulation AO_cons1} and \eqref{eqn: prob formulation AO_cons2}, respectively, one can see that problem \eqref{eqn:
DMU dual problem} can be regarded as a weighted power minimization
problem for a \emph{virtual} FD system with $K$ UMUs and
$L$ DMUs. In particular, $\{\mu_\ell\}$ and
$\{\lambda_k\}$ are the DL and UL powers in the virtual FD system, respectively, and all the variables and channels originally
associated with the DL (resp. UL) are now associated
with UL (resp. DL) in the virtual system. 

Analogous to the HD UDD which is used to develop a fixed-point method \cite{Liu1998,Wiesel2006}, we use the above FD UDD to develop a new fixed-point method to solve  {\sf (P2)}.
To see this, notice that \eqref{eqn: DMU dual
problem VDL} and \eqref{eqn: DMU dual problem VUL} must hold with
equality at the optimum. By substituting \eqref{eqn:
downlink_beam_from_dual dualADC_2} into them, constraints
\eqref{eqn: DMU dual problem VDL} and \eqref{eqn: DMU dual problem
VUL} can be equivalently written as
\begin{align}
&\lambda_k=  \Fc_k^{\rm U}(\lambdab,\mub) \triangleq  \frac{1}{\hb_k^H \Qb^{-1}(\lambdab,\mub)\hb_k/\rho_{{k}}^{\rm D}} ,~\forall k\in \Kc, \label{eq_F_U_fixed_point} \\
&\mu_{\ell}= \Fc_\ell^{\rm D}(\lambdab,\mub) \triangleq
\frac{\sum_{j=1}^L \mu_{j} \vb_\ell^H \tilde \Gb_j
\vb_\ell+\sum_{k=1}^{K}\lambda_{k} |f_{\ell k}|^2+b_\ell
 }{ |\tilde
\vb_\ell^H\gb_\ell|^2/\rho_{{\ell}}^{\rm U}}, \notag \\
&~~~~~~~~~~~~~~~~~~~~~~~~~~~~~~~~~~~~~~~~~~~~~~~~~~~~~\forall \ell\in \Lc.\label{eq_F_D_fixed_point dualADC_2}
\end{align}
By defining 
\begin{align*}
&{\bm \Fc}(\lambdab,\mub)\triangleq [\Fc_1^{\rm
	U}(\lambdab,\mub),\ldots,\Fc_K^{\rm U}(\lambdab, \mub), \notag \\
&~~~~~~~~~~~~~~~~~~~~ \Fc_1^{\rm
	D}(\lambdab,\mub), \ldots, \Fc_L^{\rm D}(\lambdab,\mub)]^T,
\end{align*} we obtain the following fixed point equation
\begin{align}\label{eqn: fix point function dualADC_2}
    [
     \lambdab^T      \mub^T
  ]^T
  = {\bm \Fc}(\lambdab,\mub).
\end{align}

\begin{Lemma}\label{lemma: fix point method}
 Suppose that problem ${\sf (P2)}$ in \eqref{eqn: prob formulation AO dualADC_2} is feasible.
 Then the optimal solution $(\lambdab^\star,\mub^\star)$ of \eqref{eqn: DMU dual problem} is the unique fixed point 
 of \eqref{eqn: fix point function dualADC_2}. Moreover, given any initial $(\lambdab^{(0)},\mub^{(0)})$, $(\lambdab^\star,\mub^\star)$ can be achieved by the following fixed-point iterations
 \begin{align}\label{eqn: fix point function dualADC_3}
 [
 (\lambdab^{(t+1)})^T      (\mub^{(t+1)})^T
 ]^T
 \leftarrow {\bm \Fc}(\lambdab^{(t)},\mub^{(t)})
 \end{align}
 as $t\to \infty$.
\end{Lemma}

{\bf Proof:}  As shown in \eqref{eqn: fix point function dualADC_2}, once {\sf (P2)} is feasible, $(\lambdab^\star,\mub^\star)$ is a fixed point.
So the fixed point exists for \eqref{eqn: fix point function dualADC_2}. To show that $(\lambdab^\star,\mub^\star)$ is the unique fixed point and the fixed-point iterations in 
\eqref{eqn: fix point function dualADC_3} converges to $(\lambdab^\star,\mub^\star)$ for arbitrary  $(\lambdab^{(0)},\mub^{(0)})$, it suffices to show that ${\bm \Fc}$ is a standard interference function 
 \cite[Theorem 2]{Yates95}; that is,
each   $\Fc_k^{\rm U}$ and $\Fc_\ell^{\rm D}$ should satisfy positivity, monotonicity
and scalability properties, for all $k\in \Kc$ and $\ell\in \Lc.$
The part of  $\Fc_k^{\rm U}$ can be proved following exactly the same arguments as in \cite[Appendix II]{Wiesel2006}, while the part of $\Fc_\ell^{\rm D}$ is easy to verity to be true.
The details are omitted. \hfill $\blacksquare$
%

Once $(\lambdab^\star,\mub^\star)$ is obtained by the above fixed-point iterations, the optimal DL beamforming direction $\{\tilde \wb_k^\star\}$ can be obtained by 
\eqref{eqn: downlink_beam_from_dual dualADC_2}. What remains for solving {\sf (P2)} is to obtain the optimal DL transmission powers $\{p_k^{\rm D\star}\}$
and UL transmission power $\{p_\ell^{\rm U\star}\}$. To show how they can be obtained, 
let us first introduce some notations. Let
$\Sb_{11}$ be a $K \times K$ matrix whose $k$th diagonal entry
is $ |\hb_k^H\tilde
\wb_k^\star|^2/\rho_k^{\rm D}-(\tilde\wb^\star_k)^H \tilde \Hb_{k} \tilde
\wb^\star_k $ and $(k,i)$th off-diagonal entry is
$-(\tilde\wb^\star_i)^H \tilde \Hb_{k} \tilde \wb^\star_i$.
Moreover, define $\Sb_{12}$ as a $K \times L$ matrix with the
$(k,\ell)$th entry being $-|f_{\ell k}|^2$. 
Similarly, let us define an $L \times L$ matrix $\Sb_{22}$ which has the $\ell$th diagonal entry being
$| \vb_\ell^H\gb_\ell|^2/\rho^{\rm U}_\ell
-\vb_\ell^H \tilde \Gb_\ell \vb_\ell$ and the
$(k,i)$th off-diagonal entry being $-\vb_\ell^H \tilde \Gb_j
\vb_\ell$. Lastly, define an $L \times K$ matrix  $\Sb_{21}$ whose $(\ell,k)$th entry is
$-\tilde \wb_k^H \LambdaB(\vb_\ell ,\Rb_{\Phib_{0}}) \tilde
\wb_k$. Then,  given $\{\wb_k^\star\}$, constraints \eqref{eqn: prob formulation AO_cons1} and \eqref{eqn: prob formulation AO_cons2}
can be compactly expressed as
\begin{align}\label{eqn: linear power sys}
 \Sb(\{\tilde \wb_k^\star\},\{\vb_\ell\})   
 \begin{bmatrix}
 \pb^{\rm D}\\
 \pb^{\rm U}
 \end{bmatrix}\succeq 
 \begin{bmatrix}
 \sigmab^2\\
 \tilde \sigma_z^2 \oneb
 \end{bmatrix},
\end{align}
where $\pb^{\rm D}=[p_1^{\rm D},\ldots,p_K^{\rm D}]^T$ and
\begin{equation} \label{eq_S_UDpower_allocation dualADC_2}
  \Sb(\{\tilde \wb_k^\star\},\{\vb_\ell\})  =
   \begin{bmatrix}
   \Sb_{11}  & \Sb_{12}\\
   \Sb_{21} & \Sb_{22}
   \end{bmatrix}.
\end{equation}
The following lemma gives the closed-form solution of optimal
power $(\{p_k^{\rm D\star}\},\{p_\ell^{\rm U\star}\})$.
\begin{Lemma}\label{lemma: primal power sol}
 Suppose that problem {\sf (P2)} in \eqref{eqn: prob formulation AO dualADC_2} is feasible, and that the optimal DL beamforming directions
 $\{\tilde \wb_k^\star\}$ are given.
 Then the optimal DL and UL powers $\pb^{\rm D\star}$ and $\pb^{\rm U\star}$ are uniquely given by
 \begin{align}\label{eqn: primal power dualADC_2}
\begin{bmatrix}
\pb^{\rm D\star}\\
\pb^{\rm U\star}
\end{bmatrix}= \Sb^{-1}(\{\tilde \wb_k^\star\},\{\vb_\ell\})   
\begin{bmatrix}
\sigmab^2\\
\tilde \sigma_z^2 \oneb
\end{bmatrix} \succ \zerob.
 \end{align}
 \end{Lemma}

{\bf Proof:} Given that {\sf (P2)} is feasible, constraints \eqref{eqn: prob formulation AO_cons1} and \eqref{eqn: prob formulation AO_cons2} will hold with equality at the optimum and 
$\pb^{\rm D\star}$ and $\pb^{\rm U\star}$ will be a solution to the linear system \eqref{eqn: linear power sys}. To show that $\pb^{\rm D\star}$ and $\pb^{\rm U\star}$ are the unique solution, one can follow a similar arguments as in \cite[Lemma 1]{SongTCOM07}. A sufficient and necessary condition for the proof to be valid is that the diagonal elements of $ \Sb_{11}$ and $ \Sb_{22}$ have to be positive, i.e.,
$ |\hb_k^H\tilde
\wb_k^\star|^2/\rho_k^{\rm D}-(\tilde\wb^\star_k)^H \tilde \Hb_{k} \tilde
\wb^\star_k >0$ for all $k\in \Kc$ and $| \vb_\ell^H\gb_\ell|^2/\rho^{\rm U}_\ell
-\vb_\ell^H \tilde \Gb_\ell \vb_\ell >0$  for all $\ell \in \Lc$. Recall from Proposition \ref{prop: FD UDD} that the optimal DL beamforming direction $\{\wb_k^\star\}$
must satisfy \eqref{eqn: DMU dual problem VDL}, i.e.,
\begin{align*}
\!\!\ 	&\lambda_k^\star|\hb_{k}^H\tilde \wb_k^\star|^2/\rho_{k}^{\rm D} -   (\tilde \wb_k^\star)^H(\lambda_{k}^\star\tilde
	\Hb_k ) \tilde
	\wb_k^\star \notag \\
\!\! &	-  { (\wb_k^\star)^H \bigg ( \sum_{i \neq k}^{K} \lambda_{i}^\star\tilde
 		\Hb_i+ \sum_{\ell=1}^L \mu_\ell^\star  \LambdaB(\vb_\ell
 		,\Rb_{\Phib_{0}}) + \Bb \bigg)
  \tilde
  \wb_k^\star } \geq 0. 
\end{align*}
Since $\sum_{i \neq k}^{K} \lambda_{i}^\star\tilde
\Hb_i+ \sum_{\ell=1}^L \mu_\ell^\star  \LambdaB(\vb_\ell
,\Rb_{\Phib_{0}}) + \Bb \succ \zerob$ (due to $\Bb\succ \zerob$), we have $ |\hb_k^H\tilde
\wb_k^\star|^2/\rho_k^{\rm D}-(\tilde\wb^\star_k)^H \tilde \Hb_{k} \tilde
\wb^\star_k >0$. Analogously, since $\{\vb_\ell\}$ satisfies \eqref{eqn: DMU dual problem VUL} and $\sum_{j\neq \ell}^L \mu_{j}^\star
\vb_\ell^H \tilde \Gb_j
\vb_\ell+\sum_{k=1}^{K}\lambda_{k}^\star |f_{\ell k}|^2+b_\ell>0$, it is clear that 
$| \vb_\ell^H\gb_\ell|^2/\rho^{\rm U}_\ell
-\vb_\ell^H \tilde \Gb_\ell \vb_\ell >0$. 
\hfill $\blacksquare$

It is worth noting that Proposition \ref{prop: FD UDD}, Lemma \ref{lemma: fix point method} and Lemma \ref{lemma: primal power sol} essentially generalize the existing UDD and fixed-point method in HD systems \cite{Liu1998,Boche2002,SongTCOM07} to the FD system. As seen, this generalization enables an efficient way to solve {\sf (P2)}.
By combing it with the subgradient method, we come up with a low-complexity algorithm to solve problem \eqref{eqn: prob formulation AO} (i.e., \eqref{eqn: power min problem dual ADC}),
which is shown in Algorithm \ref{alg: fixed-point method}. In particular, step 3 to step 7 of Algorithm \ref{alg: fixed-point method} are the fixed-point iterations to obtain $(\lambdab^\star,\mub^\star)$; step 8 and step 9 are based on \eqref{eqn: downlink_beam_from_dual dualADC_2} and Lemma \ref{lemma: primal power sol}; step 10 is the subgradient update for dealing with the ADC input power constraint \eqref{eqn: prob formulation AO_cons3}, where $s^{(r)}$ is the step size\footnote{Note that if the ADC input power constraint \eqref{eqn: prob formulation AO_cons3} is not considered, then the sbgradient loop (i.e., Steps 1, 2 and 10 to 12) can be removed; in that case, $\Bb$ and $b_\ell$'s reduce to $\Ib_{N_t}$ and $1$, respectively.}.
Comparing to directly solving the SOCP \eqref{eqn: SOCP} using a general purpose solver, the proposed fixed-point based algorithm in Algorithm \ref{alg: fixed-point method} is computationally more efficient, and therefore improving the computational efficiency of Algorithm
\ref{table: Algorithm 1}. 

\begin{algorithm}[t!]
	\caption{Proposed fixed-point based algorithm for solving
		subproblem \eqref{eqn: prob formulation AO} in Algorithm \ref{table: Algorithm 1}. } \label{alg: fixed-point method}
	\begin{algorithmic}[1]\label{table: Algorithm 2}
		\STATE {\bf Given} initial variables $\nub^{(0)}=\zerob$; set $r\leftarrow 0.$
		\REPEAT
		\STATE set $t \leftarrow 0$ and initial  $\lambdab^{(0)}=\zerob$ and $\mub^{(0)}=\zerob.$
		\REPEAT
		\STATE
		$
		[
		\ub^{(t+1)} := (\lambdab^{(t+1)})^T      (\mub^{(t+1)})^T
		]^T
		\leftarrow {\bm \Fc}(\lambdab^{(t)},\mub^{(t)})。
		$
		\STATE $t \leftarrow t+1$
		\UNTIL  $\|\ub^{(t+1)}-\ub^{(t)}\|_2/\|\ub^{(t)}\|_2 \leq \epsilon_1$. Denote the converged results as $\lambdab^\star$ and $\mub^\star$.
		\STATE Obtain $\{\tilde \wb_k^\star\}$ by \eqref{eqn: downlink_beam_from_dual dualADC_2}, and obtain $\{(p_k^{\rm D\star})\}$ and $\{(p_\ell^{\rm U\star})\}$ by \eqref{eqn: primal power dualADC_2}.
		\STATE  $\wb_k^\star\leftarrow \sqrt{(p_k^{\rm D\star})}\tilde \wb_k^\star,~\forall k\in \Kc.$
		\STATE  $\nub^{(r+1)} \leftarrow  \nub^{(r)} + s^{(r)} ( \diag ( {\rm ADC} (\Wb^\star,\pb^{\rm U\star}) -  \gamma^{\rm ADC}\Ib_{N_t}) )$. 
		\STATE $r \leftarrow r+1.$
		\UNTIL  $\|\nub^{(r+1)}-\nub^{(r)}\|_2/\|\nub^{(r)}\|_2 \leq \epsilon_2$
	\end{algorithmic}
\end{algorithm}

\section{Simulation Results}\label{sec: simulation}

 In the simulation, we consider a wireless system as described in Section \ref{subsc: signal model} and Figure \ref{fig: network diagram}. The FD BS has $10$ antennas ($N_t=10$) for simultaneous UL and DL communications \cite{Katti_ACM13,Katti_14} .
The channel coefficients $\{\hb_i\}$,$\{\gb_j\},$ $\{ f_{ji} \}$ and $\Hb_0$ are composed by large-scale path loss as well as small scale Rayleigh fadings. In particular, the path loss between the MUs and the BS is set to  $-80$ dB and that between the UMUs and DMUs is set to $-83$ dB. The SI channel has a $-10$ dB path loss \cite{Katti_ACM13,Katti_14}. Besides, following \cite{Katti_14}, there is additional $-24$ dB cross-talk path loss for neighboring antennas and further $-6$ dB path loss for farther antennas. That is, the path loss between (transmit) antenna $i$ and (receive) antenna $j$ in the SI channel is -10dB for all $i=j$ and $-34-6|i-j-1|$ dB, for all $i\neq j$, where $i,j\in \{1,\ldots,N_t\}$.
Define $\Tb$ (resp. $\Cb$) as an $N_t$ by $N_t$ Toeplitz matrix with the first row being $[1, 10^{ \frac{-24}{20} }, 10^{ \frac{-30}{20}}, \ldots,  10^{ \frac{-24-6(N_t-2)}{20}}  ]$ (resp. $[1, 0.9, 0.9^2,\ldots, 0.9^{N_t-1} $]). 
We model the correlation matrix of SI channel $\Hb_0$ as 
\begin{align}
\!\!\!\!\!   \Rb_{\Hb_0}&=  \E[{\rm vec}(\Hb_0){\rm vec}(\Hb_0)^T  ] \notag \\
   &= \bigg(10^{ \frac{-10}{10} }[{\rm vec}(\Tb){\rm vec}(\Tb)^T  ]\bigg)\odot \bigg( \oneb_{N_t \times N_t} \otimes \Cb \bigg), \label{eqn: He correlation}
\end{align}
where $\odot$ is the (element-wise) Hadamard product; $\otimes$ is the Kronecker product and $\oneb_{N_t \times N_t}$ is the $N_t$ by $N_t$ all-one matrix. The first term in the RHS of \eqref{eqn: He correlation} accounts for the cross-talk path loss, while the second term is for modeling the decreasing correlation between adjacent antennas.
To model $\Rb_{\Phib_0}$, we assume that the analog SIC scheme uses pilot-aided linear minimum mean squared error (LMMSE) channel estimation to estimate $\Hb_0$ \cite{OtterstenTSP10}. Then, $\Rb_{\Phib_0}$ is given by  
$
  \Rb_{\Phib_0} = \Rb_{\Hb_0} - \Rb_{\Hb_0} \big(  \Rb_{\Hb_0} + \frac{\sigma_z^2}{E}\Ib_{N_t^2} \big)^{-1}\Rb_{\Hb_0},
$
where $E>0$ denotes the energy of training signals. As seen, the larger $E$ is, the more powerful the analog SIC is.
If not mentioned specifically, we set various parameters as follows:
$\sigma_z^2=\sigma_i^2=-85$ dBm, $\beta_1=\beta_2=-30$ dB, $\delta_1=-50$ dB and $\delta_2=-20$ dB.
For Algorithm \ref{table: ILS for P1 bisection}, $P_{\max}$ is set to  40 dbm and $\epsilon$ is set to $10^{-3}$. Problems \eqref{eqn: F uplink} and \eqref{eqn: F downlink} are solved by the classical fixed-point method \cite{Liu1998,Wiesel2006}. For Algorithm \ref{table: Algorithm 1}, the stopping condition is set to the relative improvement of objective value of {\sf (P)} being less than $10^{-6}$.
Instead of using a convex solver to solve the SOCP \eqref{eqn: SOCP}, we use Algorithm \ref{table: Algorithm 2} to solve problem \eqref{eqn: prob formulation AO} since they two yield the same performance theoretically.
The parameters are set to $\epsilon_1=10^{-9}$, $\epsilon_2=10^{-3}$ and $s^{(r)}=1$ (constant step size). The initial $\{\vb_\ell^{(0)}\}$ is set to the zero forcing (ZF) beamfomer, i.e., $\vb_\ell^{(0)}=\Gb^{\dag}\eb_\ell$, for all $\ell \in \Lc$, where
$\Gb^\dag$ is the pseudo inverse of $\Gb=[\gb_1,\ldots,\gb_L]^T$.
Note that if the AO algorithm runs only one iteration, then it is the same as the ZF based scheme in \cite{SunTWC16}. 
Besides, if not mentioned specifically, the ADC input power constraint \eqref{eqn: main problem ADC} is not considered in order to assess how the ADC input signal power varies with system parameters if unconstrained.

\begin{figure}[t]
	\graphicspath{{fig/}} \centering 
	{\subfigure[][Feasibility rate]{\resizebox{.35\textwidth}{!}
			{\includegraphics{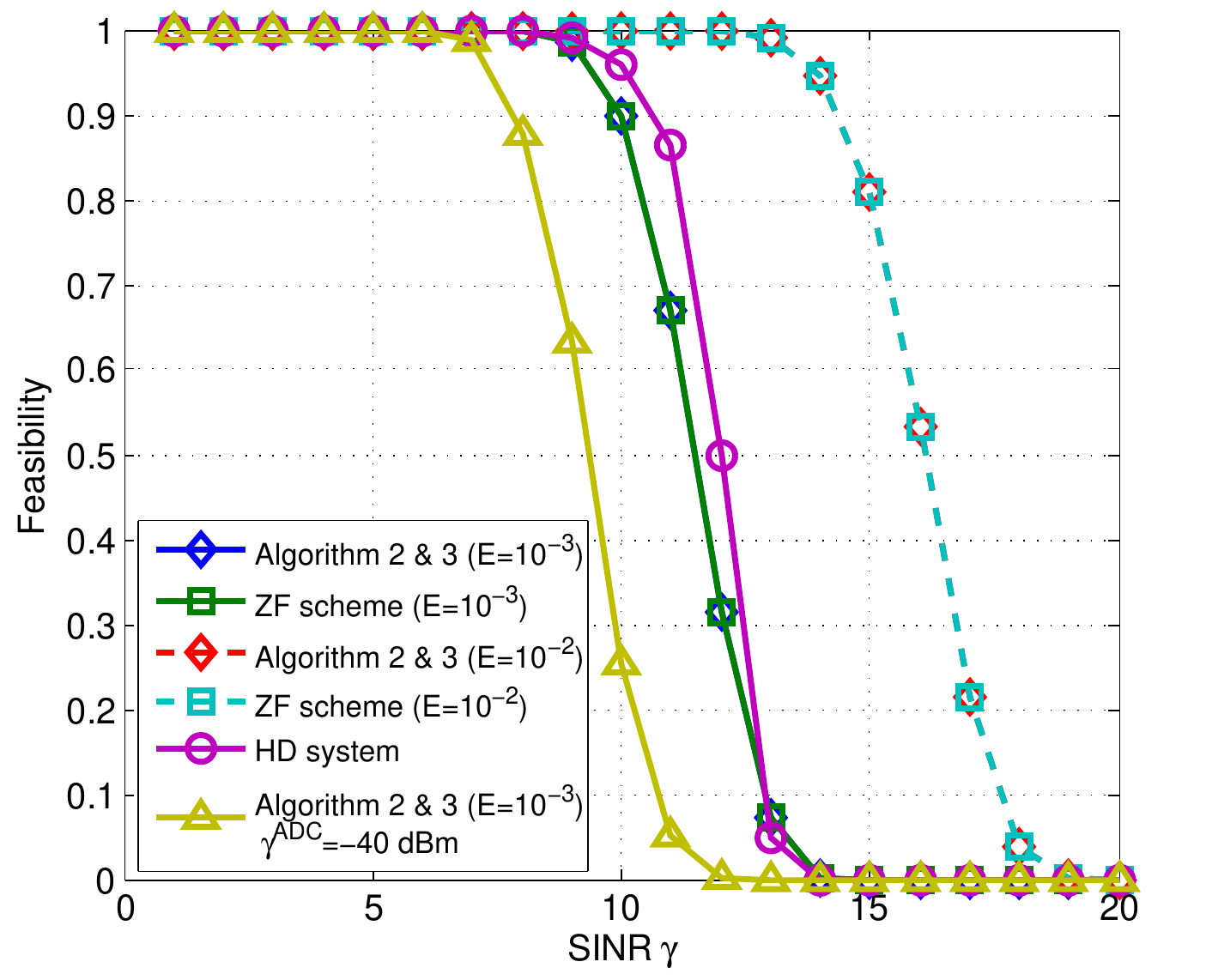}}} }
	{\subfigure[][Sum power]
		{\resizebox{.22\textwidth}{!}
			{\includegraphics{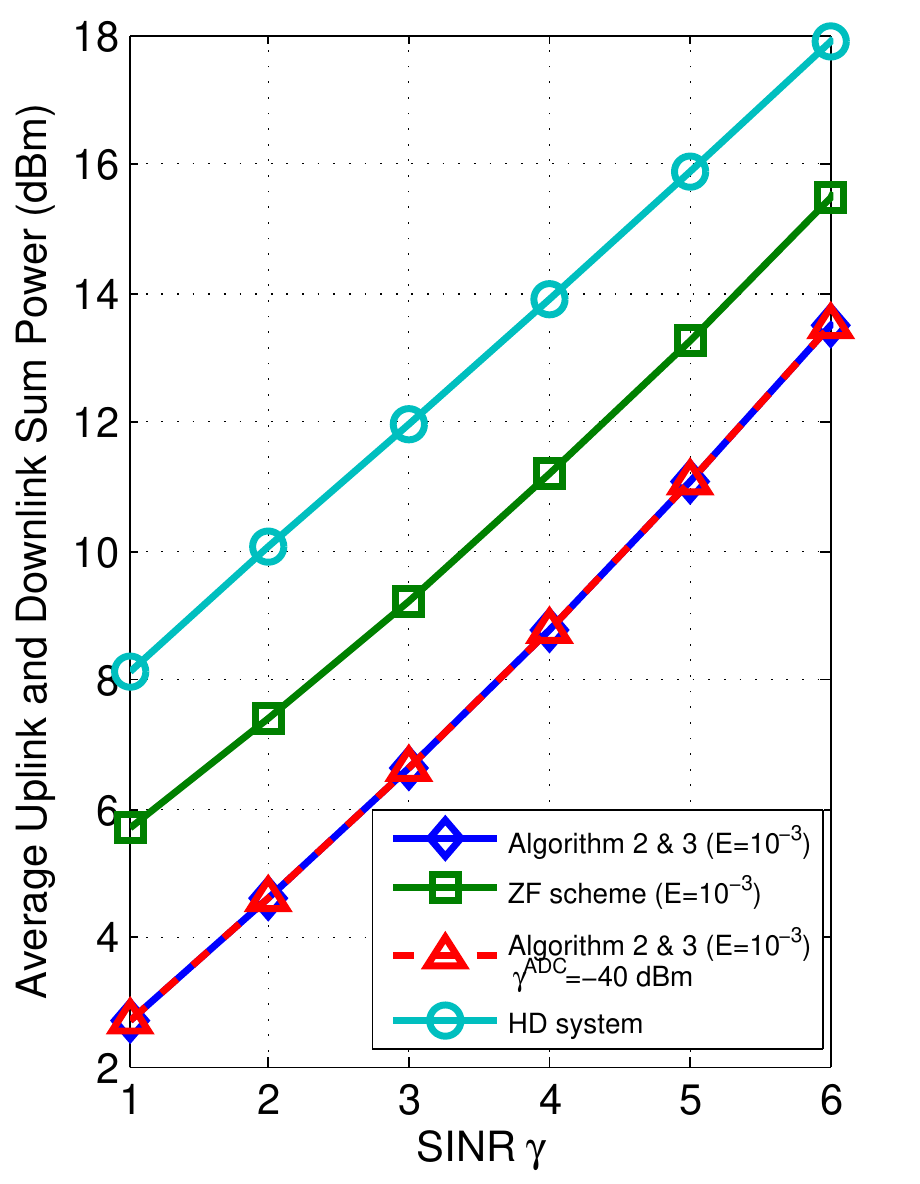}}}
	}
	{\subfigure[][ADC input signal power]
		{\resizebox{.22\textwidth}{!}
			{\includegraphics{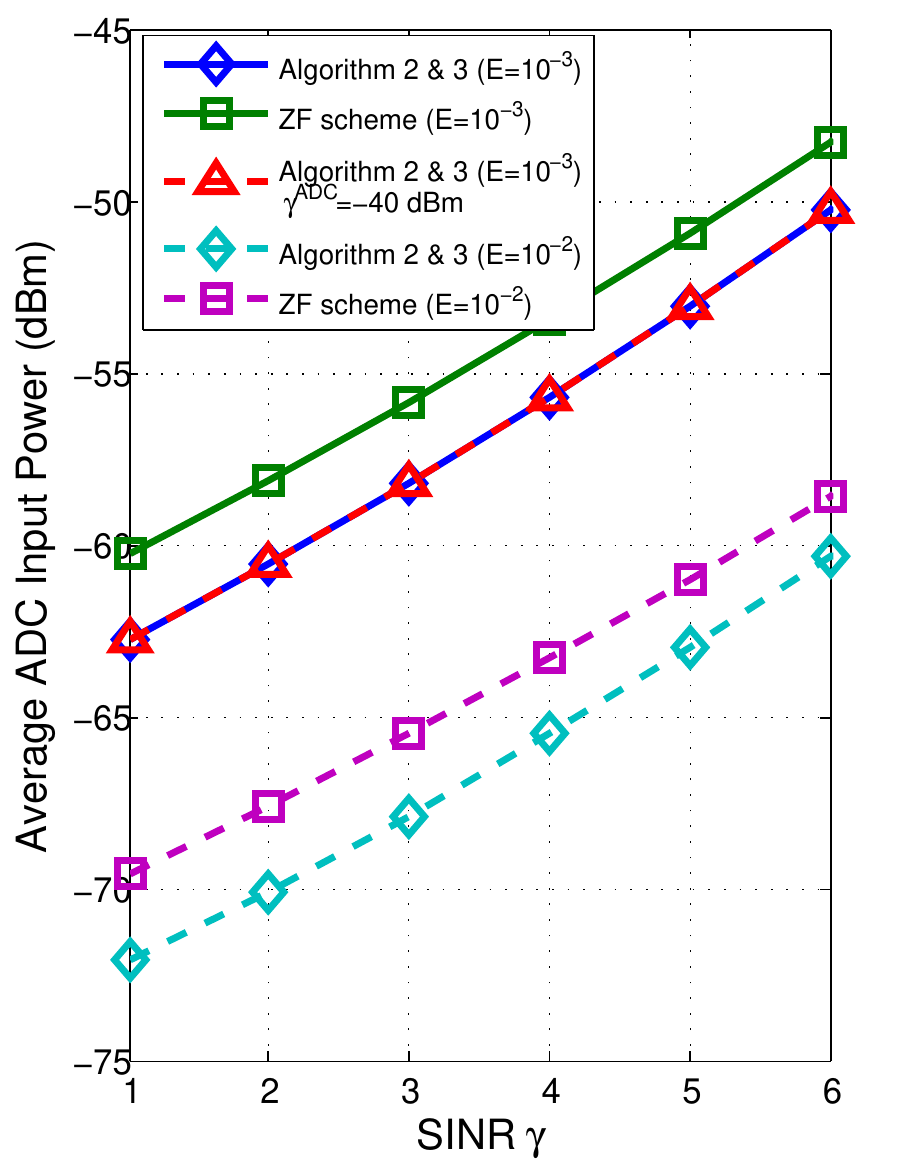}}}
	}
	\caption{Performance comparison of the proposed Algorithm 2 \& 3 and the ZF scheme; $N_t=10$, $K=L=8$.}
	\vspace{-0.5cm}\label{fig: vs sinr AO}
\end{figure}

For simplicity, we let all UMUs and DMUs have the same SINR requirement,
i.e., $\gamma\triangleq \gamma_i^{\rm D}=\gamma_\ell^{\rm U}$ for all $i\in \Kc$ and $\ell\in \Lc$.
For a fair comparison, we set the SINR of MUs in the HD system as $\gamma_{\rm HD}\triangleq 2^{2\log_2(1+\gamma)}-1$.
This implies that the information rate achieved by the HD MUs should be twice of that by the FD MUs \cite{SunTWC16}.
All simulation results are obtained by averaging over 500 channel realizations.

\begin{figure}[t]
	\graphicspath{{fig/}} \centering 
	{\subfigure[][Feasibility rate]{\resizebox{.35\textwidth}{!}
			{\includegraphics{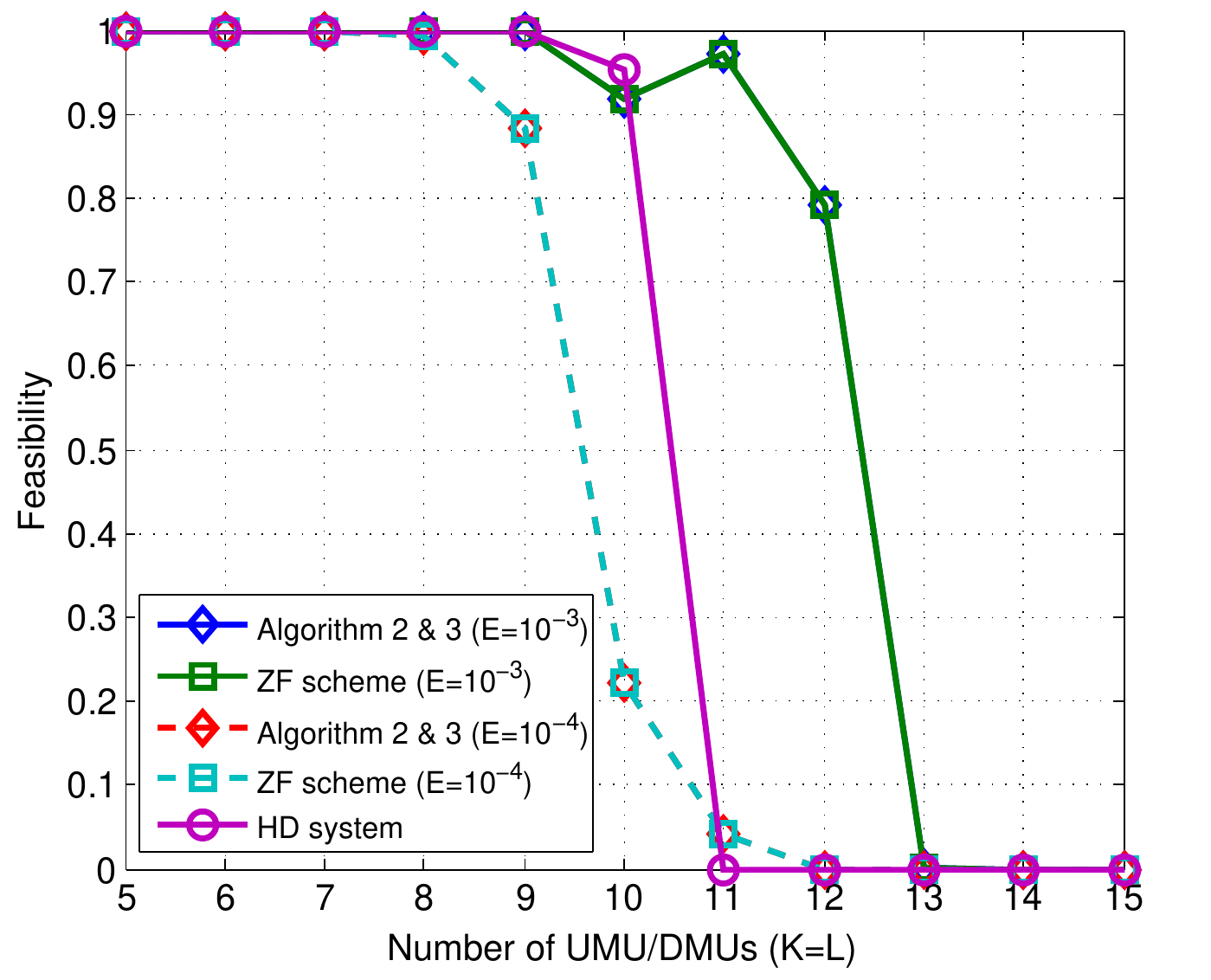}}} }
	{\subfigure[][Sum power]
		{\resizebox{.23\textwidth}{!}
			{\includegraphics{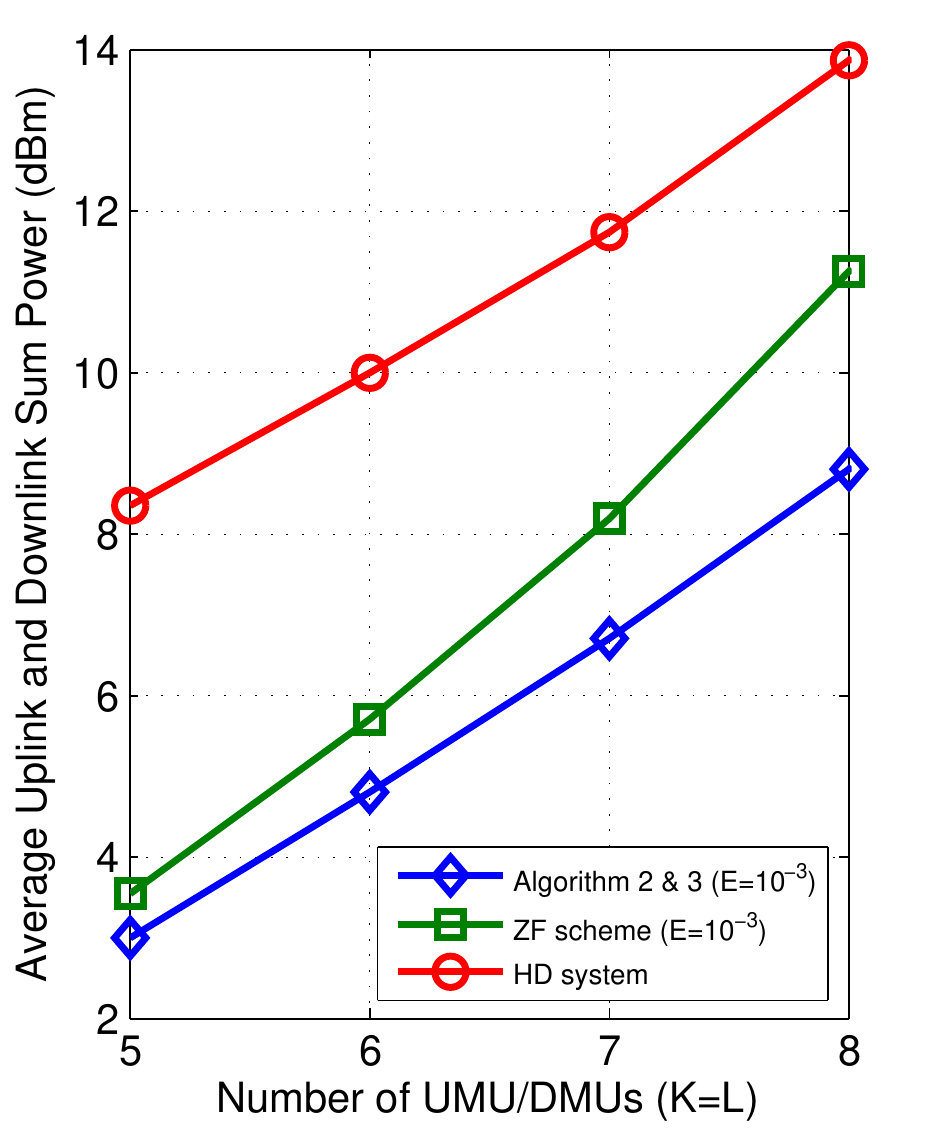}}}
	}
	{\subfigure[][ADC input signal power]
		{\resizebox{.23\textwidth}{!}
			{\includegraphics{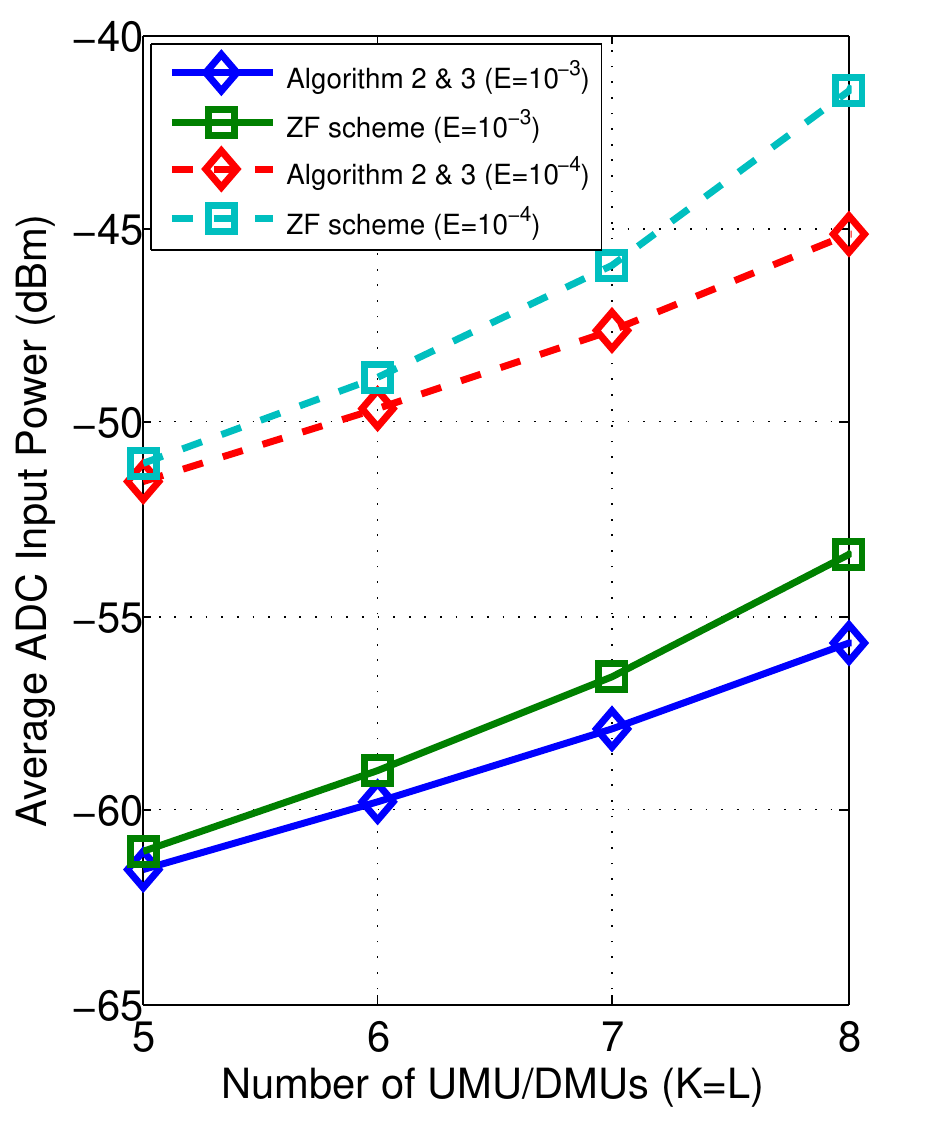}}}
	}
	\caption{Performance comparison of the proposed Algorithm 2 \& 3 and the ZF scheme; $N_t=10$, $\gamma=5$ dB.}
	\vspace{-0.4cm}\label{fig: vs user AO}
\end{figure}

\begin{figure}[t]
	\graphicspath{{fig/}} \centering 
	{\subfigure[][Feasibility rate]{\resizebox{.35\textwidth}{!}
			{\includegraphics{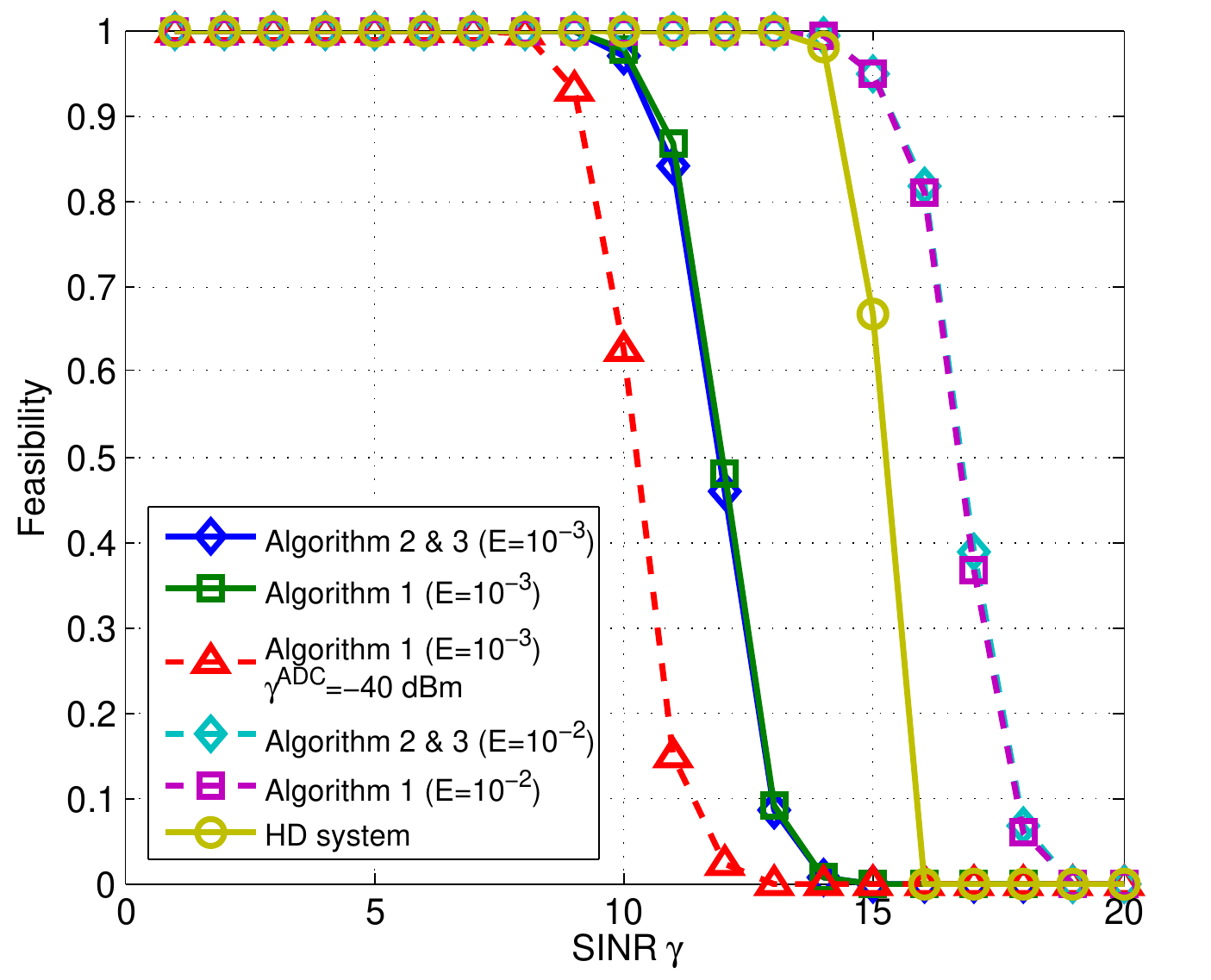}}} }
	{\subfigure[][Sum power]
		{\resizebox{.22\textwidth}{!}
			{\includegraphics{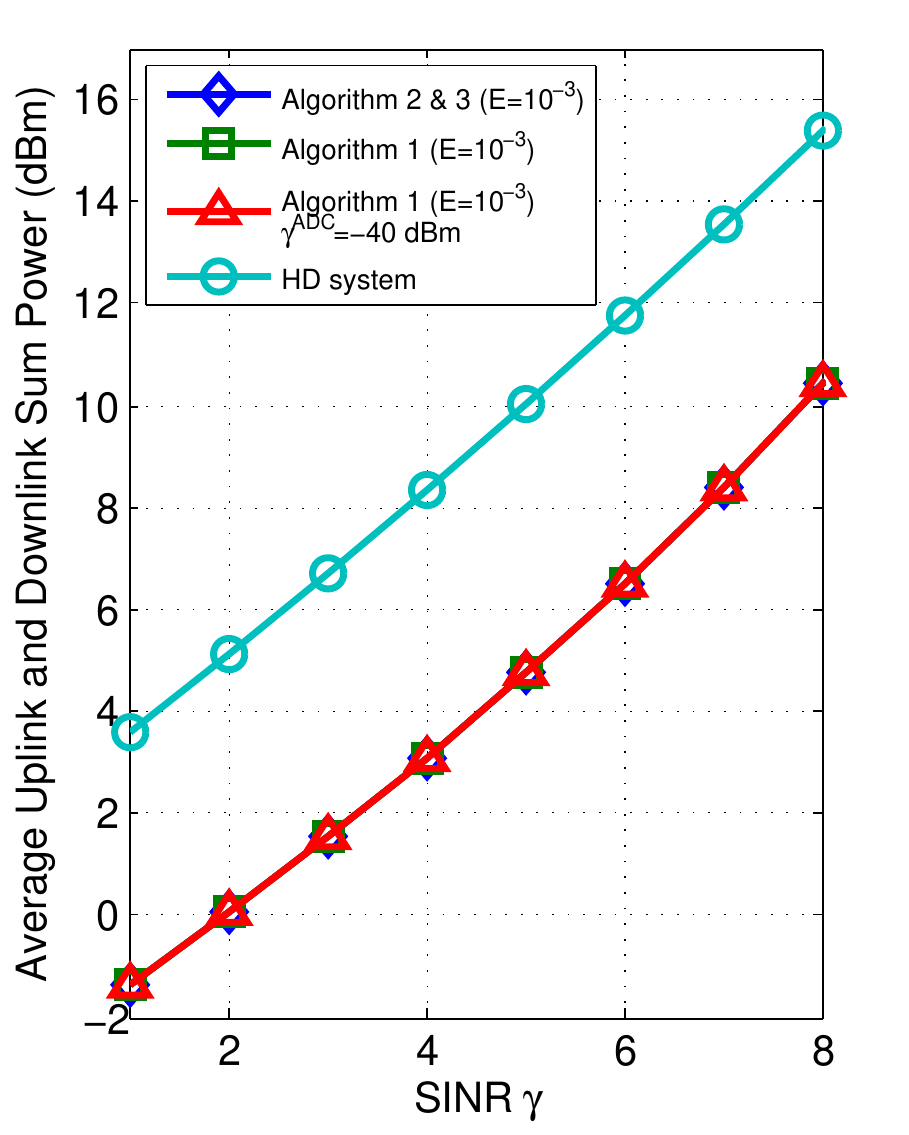}}}
	}
	{\subfigure[][ADC input signal power]
		{\resizebox{.22\textwidth}{!}
			{\includegraphics{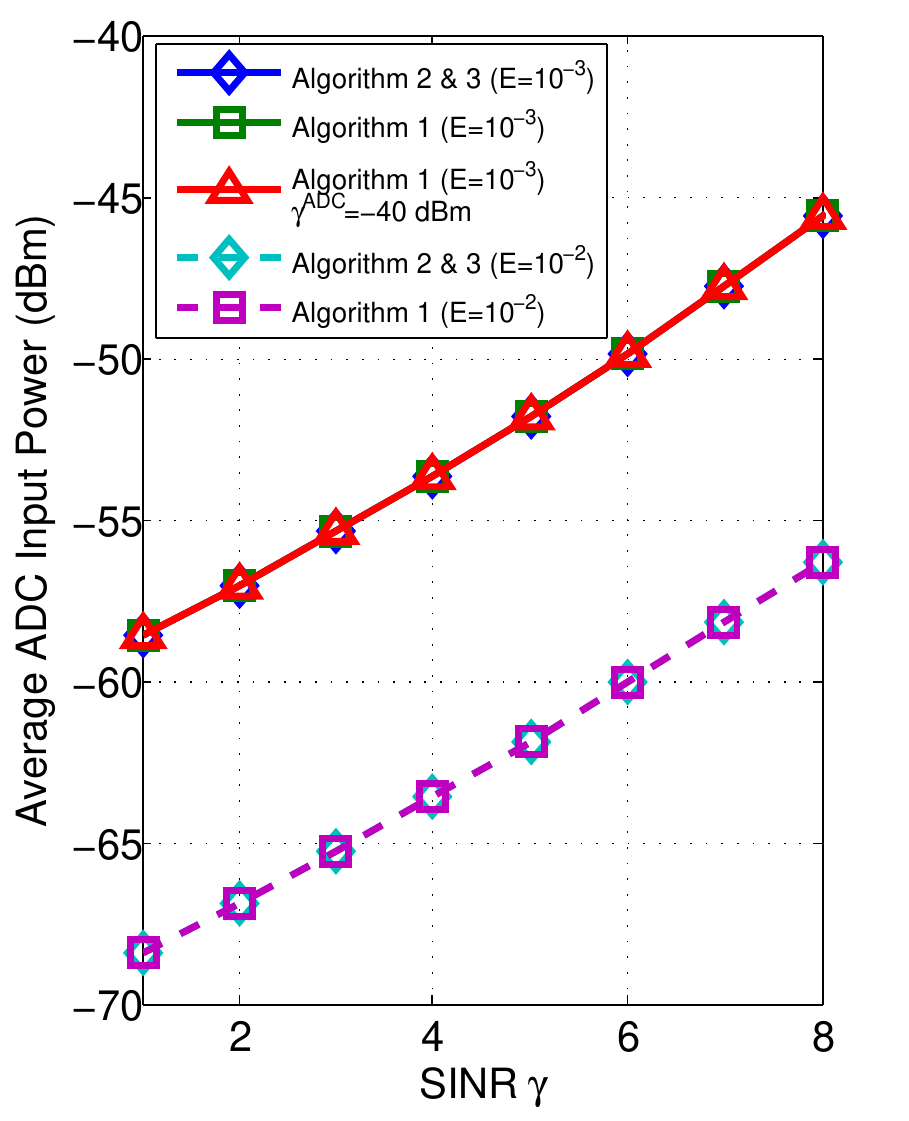}}}
	}
	\caption{Performance of the proposed Algorithm 1, 2 and 3; $N_t=10$, $K=L=5$, $\Rb_{\Phi_0}=\sigma_{\Phi_0}^2 \Ib_{N_t^2}$ where $\sigma_{\Phi_0}^2=-10$ dB.}
	\vspace{-0.5cm}\label{fig: vs sinr BS}
\end{figure}

{\bf Example 1:} In Figure \ref{fig: vs sinr AO}, we display the simulation results by comparing the proposed AO algorithm (i.e., Algorithm \ref{table: Algorithm 1} and \ref{table: Algorithm 2}) with the ZF scheme and the HD system. The number of DMUs and UMUs are eight ($K=L=8$). 
First of all, one can observe from Figure \ref{fig: vs sinr AO}(a) that the proposed AO algorithm has the same feasibility rate as the ZF scheme, which is expected, since the proposed AO algorithm is initialized by the ZF receive beamformer. However, one can see from 
Figure \ref{fig: vs sinr AO}(b) and Figure \ref{fig: vs sinr AO}(c) that\footnote{For fair comparison, in Figure \ref{fig: vs sinr AO}(b) and Figure \ref{fig: vs sinr AO}(c) , we only show the results for which all schemes under test are 100\% feasible, i.e., from $\gamma=1$ to $\gamma=6$. This principle also applies to Figure \ref{fig: vs user AO} and Figure \ref{fig: vs sinr BS}.} the AO algorithm can yield about 3 dB lower sum power and ADC input power than the ZF scheme. It can also be seen from the two figures that for both $E=10^{-3}$ and $E=10^{-2}$, the FD system using  the proposed AO algorithm is more power efficient than the HD system when both systems are feasible.

In Figure \ref{fig: vs sinr AO}, we also present the performance of the proposed AO algorithm when the ADC input power constraint \eqref{eqn: main problem ADC} is imposed with $\gamma^{\rm ADC}=-40$ dBm. One can see from Figure \ref{fig: vs sinr AO}(a) that the feasibility rate drops compared to that without the ADC input power constraint for $\gamma\geq 7$ dB. This implies that there exist realizations for which the ADC input signal power is higher than -40 dBm and this happens more frequently when $\gamma$ increases.
Since from Figure \ref{fig: vs sinr AO}(c) that the ADC input power is less than -40 dBm for $\gamma\leq 6$ dB, Figure \ref{fig: vs sinr AO}(b) shows that the AO algorithm with the ADC input power constraint performs equally well as its counterpart without the ADC input power constraint in this regime.

{\bf Example 2:}  In Figure \ref{fig: vs user AO}, we present the results by considering various numbers of DMU/UMUs in the network. 
The SINR requirement is set to $\gamma=5$ dB. As expected, the system performance (feasibility rate and sum power) deteriorates when the number of MUs in the network increases.
The ADC input power also increases since the SI power can increase both with the number of UMUs and the number of DMUs.
One may notice that the feasibility rate of the proposed AO algorithm ($E=10^{-3}$) oscillates between $K=L=10$ and $K=L=11$. This is because the AO algorithm is initialized by different 
$\{\vb_\ell^{(0)}\}$ when the number of UMUs changes. Finally, it can be observed from Figure \ref{fig: vs user AO}(b) and Figure \ref{fig: vs user AO}(c) that the more number of MUs in the network, the better the AO algorithm performs than the ZF scheme.

{\bf Example 3:} In the last example, let us examine the performance of Algorithm \ref{table: ILS for P1 bisection} by assuming that the SI channel errors are i.i.d. Specifically, we let
$\Rb_{\Hb_0}=\sigma_{\Hb_0}^2 \Ib_{N_t^2}$ where $\sigma_{\Hb_0}^2=-10$ dB\footnote{So $\Rb_{\Phi_0}=\sigma_{\Phi_0}^2 \Ib_{N_t^2}$ where $\sigma_{\Phi_0}^2$ is approximately -95 dB when $E=10^{-3}$ and  -105 dB when $E=10^{-2}$.}, and let $K=L=5$.
The simulation results are presented in Figure \ref{fig: vs sinr BS}. 
Firstly, one can observe from Figure \ref{fig: vs sinr BS}(a) that there exists small discrepancy between the feasibility rates of the AO algorithm (Algorithm \ref{table: Algorithm 1} and \ref{table: Algorithm 2}) and the bisection algorithm (Algorithm \ref{table: ILS for P1 bisection}), especially when the feasibility rates are not 100\%. We suspect that the AO algorithm actually achieves the same solution as the bisection algorithm under the simulation setting, and the discrepancy in the feasibility rate are caused by some numerical issues. This is evidenced by observing from Figure \ref{fig: vs sinr BS}(b) and Figure \ref{fig: vs sinr BS}(c) that the AO and bisection algorithms essentially yield the same sum power and ADC input power when both methods are feasible.
Therefore, the simulation results imply that the AO algorithm may have achieved optimal or near-optimal solutions for the cases with $\Rb_{\Hb_0}=\sigma_{\Hb_0}^2 \Ib_{N_t^2}$.
From Figure \ref{fig: vs sinr BS}(a), we again see that with the ADC input power constraint, the feasibility rates of the considered design problem decreases.

\section{Conclusion}\label{sec: conclusion}
In this paper, by taking into account the non-ideal RF chains and analog/digital SIC, we have formulated the QoS-based linear transceiver design problem {\sf (P)} which is not only  constrained by the minimum SINR requirement of MUs but also by the maximum power at the ADC input.
While such problem is non-convex and difficult to solve in general, we have shown that it can be solved to the global optimum when the SI channel estimation errors are i.i.d.
Specifically, we have developed a bisection algorithm (Algorithm \ref{table: ILS for P1 bisection})  that can achieve the global optimal solution in a polynomial-time complexity.
To handle  {\sf (P)} in general, we have proposed a suboptimal AO algorithm (Algorithm \ref{table: Algorithm 1}). Moreover, we have generalized the UDD in the HD system to the FD system and proposed an efficient fixed-point based algorithm for solving \eqref{eqn: prob formulation AO} (Algorithm \ref{table: Algorithm 2}).
The presented simulation results have shown that the proposed AO algorithm outperforms the ZF scheme. Moreover, when the SI channel estimation errors are i.i.d., the simulation results also suggest that the AO algorithm achieves a near-optimal solution. It is also observed that the FD system is more power efficient than the HD system especially when the QoS requirements 
are less stringent or when the number of MUs is moderate.

	\vspace{-0.0cm}
\appendices {\setcounter{equation}{0}
	\renewcommand{\theequation}{A.\arabic{equation}}
	
	%
	%
\section{Derivation of \eqref{eqn: omega0} and \eqref{eqn: lambda0} }\label{appx: derivations}
The derivation involves basic vector and matrix algebra. Note that we can write the first term of in the RHS of \eqref{eqn: Sigma W} as
\begin{align}
&\delta_1 \vb^H_\ell\E[\Phib_{0}\xb(t)\xb^H(t)\Phib_0^H]\vb_\ell \notag \\
&~~=\delta_1\vb^H_\ell\E\bigg[\Phib_{0}\Wb\Wb^H\Phib_0^H\bigg]\vb_\ell  \label{eqn: sinr term 11} \\
&~~=\delta_1 \sum_{k=1}^K\wb_k^H \E\bigg[\Phib_{0}^H \vb_\ell\vb_\ell^H\Phib_0\bigg]\wb_k.  \label{eqn: sinr term 12}
\end{align}
Besides, by \eqref{eqn: u_tx}, we can write the second term in the RHS of \eqref{eqn: Sigma W} as
\begin{align}
&\delta_2\vb_\ell^H  \E[\Phib_{0}\ub_{\rm tx}[n]\ub_{\rm tx}^H[n]\Phib_0^H] \vb_\ell \notag \\
&~~=(\delta_2\beta_1) \vb_\ell^H  \E[\Phib_{0} \diag(\Wb\Wb^H) \Phib_{0}^H ] \vb_\ell \label{eqn: sinr term 21} 
\\
\!\!\!\!&~~=(\delta_2\beta_1) \sum_{k=1}^K  \sum_{n=1}^{N_t}   (\vb_\ell^H  \E[\Phib_{0} \eb_n\eb_n^T \Phib_{0}^H ] \vb_\ell)  |w_{kn}|^2  \notag 
\\
\!\!\!\!&~~= (\delta_2\beta_1)  \sum_{k=1}^K\wb_k^H    \diag\bigg( \{ \vb_\ell^H \E[\Phib_{0} \eb_n\eb_n^T \Phib_{0}^H ] \vb_\ell\}_n \bigg)
\wb_k.  \label{eqn: sinr term 22}
\end{align}
Analogously, by \eqref{eqn: BS receive signal2} and \eqref{eqn: urx}, the third term in the RHS of \eqref{eqn: Sigma W} can be written as
\begin{align}
&\delta_2 \vb_\ell^H \E[\ub_{\rm rx}[n]\ub_{\rm rx}^H[n]] \vb_\ell \notag \\
&=
(\delta_2  \beta_2)  \vb_\ell ^H \diag\bigg( \sum_{j=1}^{L} \gb_{j}\gb_{j}^H p^{\rm
	U}_j + \E[\Phib_{0}\Wb\Wb^H\Phib_{0}^H] +
\beta_1 \E[\Phib_{0}\diag(\Wb\Wb^H)\Phib_{0}^H] +\sigma_z^2\Ib_{N_t} \bigg)\vb_\ell
\notag \\
&=  (\delta_2  \beta_2)  \vb_\ell ^H \bigg( \sum_{j=1}^{L}  p^{\rm
	U}_j \diag( |\gb_{j}|^2) \bigg)\vb_\ell
+ (\delta_2  \beta_2)  \vb_\ell ^H \diag\bigg(  \E[\Phib_{0}\Wb\Wb^H\Phib_{0}^H]
\notag \\
&~~~~+
\beta_1 \E[\Phib_{0}\diag(\Wb\Wb^H)\Phib_{0}^H] +\sigma_z^2\Ib_{N_t} \bigg)\vb_\ell  \label{eqn: sinr term 31}  \\
&=  (\delta_2  \beta_2) \bigg[ \vb_\ell ^H \bigg( \sum_{j=1}^{L}  p^{\rm
	U}_j \diag( |\gb_{j}|^2) \bigg)\vb_\ell +   \sum_{k=1}^K \wb_k^H \bigg( \sum_{n=1}^{N_t} |v_{\ell,n}|^2
\E\{\Phib_0^H \eb_n \eb_n^T\Phib_0\}\bigg) \wb_k \notag \\
&+
 \beta_1  \sum_{k=1}^K \wb_k^H  \diag\bigg( \!\!\bigg\{\!\! \vb_\ell^H \diag\bigg(\E[\Phib_{0} \eb_n\eb_n^T \Phib_{0}^H ]\bigg) \vb_\ell\bigg\}_n \bigg)  \wb_k  + \sigma_z^2 \|\vb_\ell\|_2^2 \bigg].     \label{eqn: sinr term 32}
\end{align}
To simplify the notations, define
\begin{align}\label{eqn: Omega}
&\Omegab(\Wb, \Rb_{\Phib_0})
\triangleq \delta_1 \E\bigg[\Phib_{0}\Wb\Wb^H\Phib_0^H\bigg]
+ (\delta_2\beta_1) \E[\Phib_{0} \diag(\Wb\Wb^H \Phib_{0}^H ] \notag \\
&~~~~+(\delta_2  \beta_2)  \diag\bigg(  \E[\Phib_{0}\Wb\Wb^H\Phib_{0}^H]
+
\beta_1 \E[\Phib_{0}\diag(\Wb\Wb^H)\Phib_{0}^H]+\sigma_z^2\Ib_{N_t} \bigg),
 \\
&\LambdaB(\vb_\ell, \Rb_{\Phib_0})
\triangleq \delta_1 \E\{\Phib_0^H \vb_\ell\vb_\ell^H \Phib_0 \}
+  \delta_2\beta_2 \sum_{n=1}^{N_t} |v_{\ell,n}|^2 \bar \Rb_{\Phi_0,n}
+  \delta_2\beta_1 \diag(\{ \vb_\ell^H \widetilde \Rb_{\Phi_0,n} \vb_\ell\}_n),
\label{eqn: LambdaB}
\end{align}
where 
\begin{align}
\!\!&\Rb_{\Phi_0,n} \triangleq \E\{ \Phib_0 \eb_n \eb_n^T  \Phib_0^H\},
\bar \Rb_{\Phi_0,n} \triangleq \E\{\Phib_0^H \eb_n \eb_n^T \Phib_0\}, \label{eqn: Rn}\\
& \widetilde \Rb_{\Phi_0,n} \triangleq \Rb_{\Phi_0,n} + \beta_2 \diag(\Rb_{\Phi_0,n}), 
\end{align}
and $ \tilde \Gb_j \triangleq \gb_j\gb_j^H +  \delta_2\beta_2 \diag(|\gb_j|^2)$ and $\tilde \sigma_z^2 \triangleq (1+ \delta_2\beta_2 )\sigma_z^2.$
Then, \eqref{eqn: omega0} and \eqref{eqn: lambda0} can be obtained from \eqref{eqn: sinr term 11} to \eqref{eqn: LambdaB}.

\section{Proof of Proposition \ref{property: eta and P_eta}}
\label{appx: proof of prop 2}
To show part (a), suppose that {\sf (P1)} has two sets of solutions with corresponding $\{\wb_k,p_\ell^{\rm U}\}$ denoted by $\{\hat \wb_k\, \hat p_\ell^{\rm U} \}$ and $\{\tilde \wb_k, \tilde p_\ell^{\rm U}\}$.
Let $\hat \eta = \sum_{k=1}^K \|\hat \wb_k\|_2^2$ and $\tilde \eta = \sum_{k=1}^K \|\tilde \wb_k\|_2^2$.
Then it should be
\begin{align}
\hat \eta + \sum_{\ell=1}^{L} \hat p_\ell^{\rm U} = \tilde \eta+ \sum_{\ell=1}^{L} \tilde p_\ell^{\rm U}
\end{align}
Without loss of generality, suppose that $\hat \eta < \tilde \eta.$ Then $ \sum_{\ell=1}^{L} \hat p_\ell^{\rm U}>\sum_{\ell=1}^{L} \tilde p_\ell^{\rm U}$.
According to the proof of Proposition \ref{prop: two stage}, given a value of $\sum_{k=1}^K\|\wb_k\|_2^2=\eta$, the optimal $\{ p_\ell^{\rm U}\}$ must satisfy \eqref{eqn: uplink sol} and is unique.
As a result, for $\hat \eta < \tilde \eta$, we must have  $ \sum_{\ell=1}^{L} \hat p_\ell^{\rm U} < \sum_{\ell=1}^{L} \tilde p_\ell^{\rm U}$ which is a contradiction. Therefore, we obtain $\hat \eta = \tilde \eta.$

To show part (b), we observe that when $\eta=\eta^\star=\sum_{k=1}^K\|\wb_k^\star\|_2^2$, the UL powers $\{p_\ell^{\rm U}\}$ of both {\sf (P1)} and  {\sf (P$_\eta$)}
are uniquely determined by the system equations in \eqref{eqn: uplink sol}. Therefore, by Lemma \ref{lemma: HDU sol}, given $\eta=\eta^\star$, {\sf (P$_\eta$)} has the same set of optimal UL powers
and UL beamformers as {\sf (P1)},
which we denote as $\{p_\ell^{\rm U \star}\}$ and $\{\vb_\ell^{\star}\}$, respectively. With $\{p_\ell^{\rm U},\vb_\ell\}$ fixed by $\{p_\ell^{\rm U \star},\vb_\ell^{\star}\}$  in {\sf (P1)} and  {\sf (P$_\eta$)},
one can show that he optimal DL beamformers $\{\wb_k\}$ for both problems {\sf (P1)} and  {\sf (P$_\eta$)} must also be solutions to the following problem
\begin{subequations}\label{eqn: F downlink 2}
	\begin{align}
	\min_{ \substack{\{\wb_k\} }}~&\sum_{k=1}^{K}
	\|\wb_k\|_2^2  \\
	{\rm s.t.}~~&
	\frac{|\hb_{i}^H\wb_{i}|^2/\rho_i^{\rm D}}
	{ \sum_{k=1}^K \wb_k^H \tilde \Hb_i \wb_k +
			\hat \sigma_i^2(\pb^{\rm U\star})}\geq 1,~i\in \Kc, \\
	&  \sum_{k=1}^K\|\wb_k\|_2^2 \leq \eta^\star, \\
	&\overline {\rm ADC} (\Wb,\pb^{\rm U\star})  \preceq  \gamma^{\rm ADC}\Ib_{N_t}. 
	\end{align}
\end{subequations}
It is not difficult to verify that problem \eqref{eqn: F downlink 2} has a unique solution up to a phase shift (e.g., see \cite[Lemma 4]{ShiQJ16}).
So $\sum_{k=1}^K \|\wb_k(\eta)\|_2^2=\sum_{k=1}^K \|\wb_k^\star\|_2^2=\eta^\star$ and therefore
$\{\wb_k(\eta^\star)\}$ of {\sf (P$_\eta$)} is also optimal to {\sf (P1)}.

Part (c) is true as $\{p_\ell^{\rm U}(\eta)\}$ increases with $\eta$ and consequently $\sum_{k=1}^K \|\wb_k(\eta)\|_2^2$ is increasing  with $\eta$.
\hfill $\blacksquare$

\section{Proof of Proposition \ref{prop: eta nd P_eta 2}}\label{appx: proof of prop 3}
{\bf Proof of part (a):}  
It is easy to see that $\sum_{k=1}^K \|\wb_k(\eta)\|_2^2$ is increasing with $\eta$. To show the concavity, we first prove that the CCI $\sum_{j=1}^L p_{j}^{\rm U}(\eta) |f_{j i}|^2$ in \eqref{eqn: F downlink C1} is a concave function of $\eta$ for any $i \in \Kc $.  Recall Proposition \ref{prop: two stage} that the optimal $\{ p_{\ell}^{\rm U}(\eta) \} $ can be obtained by solving \eqref{eqn: F uplink}.
Alternatively, let us consider the following problem
\begin{subequations}\label{eqn: F uplink arbitrary weights}
	\begin{align}
	\min_{ \substack{\{ \vb_\ell\},  \{p_\ell^{\rm U}\geq 0\}}}~& \sum_{\ell=1}^L a_\ell p_\ell^{\rm U} \\
	{\rm s.t.}~~
	&   \frac{p_\ell^{\rm U}| \vb_\ell^H\gb_\ell|^2 /\rho_\ell^{\rm U}}
	{\sum_{j =1 }^L p_j^{\rm U}  \vb_\ell^H \tilde \Gb_j \vb_\ell +
		(\xi\eta + \tilde \sigma_z^2)}  \geq 1,~\ell\in \Lc, \label{eqn: F uplink arbitrary weights C2}   \\
	& \|\vb_\ell\|_2=1.~\ell\in \Lc.
	\end{align}
\end{subequations}
where $a_\ell> 0$, $\ell=1,\ldots,L,$ are some weighting coefficients. Since at the optimum the constraint \eqref{eqn: F uplink arbitrary weights C2} holds with equality, the optimal $\{ p_{\ell}^{\rm U} \} $ of problem \eqref{eqn: F uplink arbitrary weights} also satisfies \eqref{eqn: uplink sol}.
As \eqref{eqn: uplink sol} admits only a unique solution, $\{  p_{\ell}^{\rm U}(\eta) \}$ of  \eqref{eqn: F uplink} is also the optimal solution to problem  \eqref{eqn: F uplink arbitrary weights}.
By this fact and by applying Lemma \ref{lemma: HDU SDP} to \eqref{eqn: F uplink arbitrary weights}, we obtain
\begin{subequations}\label{eqn: HDU SDP v2}
	\begin{align}
	&\sum_{\ell=1}^L a_\ell p_\ell^{\rm U}(\eta) =  \max_{  \{p_\ell^{\rm U}\geq 0\} }~
	\sum_{\ell=1}^L a_\ell p_\ell^{\rm U} \\
	&{\rm s.t.}~~
	\sum_{j =1 }^L p_j^{\rm U}  \tilde \Gb_j
	+ (\xi\eta+ \tilde \sigma_z^2)\Ib_{N_t} \succeq
	\frac{p_\ell^{\rm U}}{\rho_\ell^{\rm U}}\gb_\ell\gb_\ell^H,~\forall \ell\in \Lc.
	\label{eqn: HDU SDP v2 C1}
	\end{align}
\end{subequations}
Define an indicator function as
\begin{align}
\Ic(\pb^{\rm U},\eta)=\bigg\{
\begin{array}{ll}
0  &{\rm if~}  (\pb^{\rm U},\eta) ~{\rm satisfies~}\eqref{eqn: HDU SDP v2 C1},\\
\infty    & {\rm otherwise}.
\end{array}
\end{align} Thus one can write \eqref{eqn: HDU SDP v2} as
\begin{align}\label{eqn: F uplink SDP 2}
\sum_{\ell=1}^L a_\ell p_\ell^{\rm U}(\eta)=   \max_{ \substack{ \{p_\ell^{\rm U}\geq 0\}}}~& \sum_{\ell=1}^L a_\ell p_\ell^{\rm U} - \Ic(\pb^{\rm U},\eta).
\end{align}
Note that $\Ic(\pb^{\rm U},\eta)$ is jointly convex w.r.t. $\eta$ and $\pb^{\rm U}$.
So, by applying the maximization property of concave functions \cite[Chapter 3]{BK:BoydV04} to \eqref{eqn: F uplink SDP 2}, we obtain that $\sum_{\ell=1}^L a_\ell p_\ell^{\rm U}(\eta)$  is a concave function of $\eta$.
By letting $a_\ell=|f_{\ell i}|^2$, we obtain that $\sum_{j=1}^L  p_j^{\rm U}(\eta) |f_{ji}|^2$ in \eqref{eqn: F downlink C1}  is a concave function of  $\eta$.

We use the concavity of $\sum_{j=1}^L  p_j^{\rm U}(\eta) |f_{ji}|^2$  to show that $\sum_{k=1}^K \|\wb_k(\eta)\|_2^2$ of {\sf (P$_\eta$)}  
is also concave. 
Firstly, by Proposition \ref{prop: two stage}, $\{\wb_k(\eta)\}$ can be obtained by solving \eqref{eqn: F downlink}.
Since ADC power constraint \eqref{eqn: F downlink C2}  can be explicitly written as
\begin{align}\label{eqn: ADC explicit}
\sum_{k=1}^K \|\wb_k\|_2^2 \leq  \frac{ \gamma^{\rm ADC}　-　 \sum_{j=1}^L p_j^{\rm U}(\eta) |\eb_n^T\gb_j|^2 - \sigma_z^2}{\sigma_{\Phi_0}^2(1+\beta_1)}
\end{align}
for $~n=1,\ldots,N_t$, problem \eqref{eqn: F downlink} can be solved by first solving 
\begin{subequations}\label{eqn: F downlink v2}
	\begin{align}
	 \min_{ \substack{\{\wb_k\} }}~&\sum_{k=1}^{K}
	\|\wb_k\|_2^2  \\
	{\rm s.t.}~~&
	\frac{|\hb_{i}^H\wb_{i}|^2/\rho_i^{\rm D}}
	{ \sum_{k=1}^K \wb_k^H \tilde \Hb_i \wb_k +
		\hat \sigma_i^2(\pb^{\rm U}(\eta))}\geq 1,~i\in \Kc,
	\end{align}
\end{subequations}
followed by checking whether the optimal $\{\wb_k\}$ satisfies \eqref{eqn: ADC explicit} or not.
So given that {\sf (P$_\eta$)} is feasible, the optimal value of \eqref{eqn: F downlink v2} is 
$\sum_{k=1}^K\|\wb_k(\eta)\|_2^2$.
By applying Lemma \ref{lemma: HDD sol}, \eqref{eqn: F downlink v2} has a virtual UL problem 
\begin{subequations}\label{eqn: HDD virtual uplink}
	\begin{align}
	&\sum_{k=1}^K\|\wb_k(\eta)\|_2^2=   \min_{ \{ \tilde \wb_i\},  \{\lambda_i\geq 0\} }~
	\sum_{i=1}^K \lambda_i 	\hat \sigma_i^2(\pb^{\rm U}(\eta))   \label{eqn: HDD virtual uplink C0} \\
	&~~~~{\rm s.t.}~~
	\frac{\lambda_i| \tilde \wb_i^H\hb_i|^2 /\rho_i^{\rm D}}
	{\sum_{k =1 }^K \lambda_k \tilde \wb_i^H \tilde \Hb_k \tilde \wb_i
		+ \|\tilde \wb_i\|_2^2}  \geq 1,  \label{eqn: power min problem HD ul C1} \\
	&~~~~~~~~~~ \|\tilde \wb_i\|_2=1.~i \in \Kc,
	\end{align}
\end{subequations}
which has the same optimal value as \eqref{eqn: F downlink v2}.
Notice that, similar to Lemma \ref{lemma: HDU sol}, the optimal $\{ \lambda_i \}$ of \eqref{eqn: HDD virtual uplink} is uniquely determined by equations
\begin{align}\label{eqn: classical uplink fixed point eqs lambda}
\frac{\lambda_i}{\rho_i^{\rm D}} \hb_i^H\bigg( \sum_{k =1 }^K \lambda_k \tilde \Hb_k
+ \Ib_{N_t} \bigg)^{-1}\hb_i =1,~\forall i \in \Kc.
\end{align}
Note that \eqref{eqn: classical uplink fixed point eqs lambda} is independent of $\eta$, and therefore the optimal $\{\lambda_i\}$
of \eqref{eqn: HDD virtual uplink} is a constant w.r.t. $\eta$.
Since $\sum_{j=1}^L  p_j^{\rm U}(\eta) |f_{ji}|^2$  is a concave function of  $\eta$, we conclude that the optimal objective value $\sum_{i=1}^K \lambda_i \big (\sum_{j=1}^{L} p_j^{\rm U}(\eta) |f_{ji}|^2 +
\sigma_i^2 \big) $ of \eqref{eqn: HDD virtual uplink}, which is equal to  $    \sum_{k=1}^{K}
\|\wb_k(\eta)\|_2^2$, is a concave function of $\eta$. The proof is thus complete.

{\bf Proof of part (b):} To show sufficiency of part (b), 
by Proposition \ref{property: eta and P_eta}(c), $F(\eta)=\sum_{k=1}^K \|\wb_k(\eta)\|_2^2+ \sum_{\ell}^{L} p_\ell^{\rm U}(\eta) < F(\eta^\star)$
for $\eta < \eta^\star$. Suppose that $\sum_{k=1}^K \|\wb_k(\eta)\|_2^2 \leq \eta$.
Then $\{\wb_k(\eta),\vb_\ell(\eta),p_\ell^{\rm U}(\eta)\}$ of  {\sf (P$_\eta$)} is also a feasible solution to
{\sf (P1)}. It implies $F(\eta^\star)\leq F(\eta)$ which however is a contradiction. So the sufficiency of part (b) is true.

We draw the function $y=\sum_{k=1}^K \|\wb_k(\eta)\|_2^2$ and $y=\eta$ in Figure \ref{fig: y}.
Specifically, note that $\sum_{k=1}^K \|\wb_k(0)\|_2^2>0$, and by Proposition \ref{property: eta and P_eta}(b), the function $y=\sum_{k=1}^K \|\wb_k(\eta)\|_2^2$ intersects with the line $y=\eta$ at $\eta^\star$. 
Moreover, by part (a), $y=\sum_{k=1}^K \|\wb_k(\eta)\|_2^2$ is concave and increasing. 
Therefore, $y=\sum_{k=1}^K \|\wb_k(\eta)\|_2^2$
must be below $y=\eta$ when $\eta >\eta^\star$; that is, $\sum_{k=1}^K \|\wb_k(\eta)\|_2^2 < \eta$ for $\eta >\eta^\star$. 
So the necessity of part (b) is true.

Finally, let us show that {\sf (P1)} is infeasible if and only if {\sf (P$_\eta$)} has $\sum_{k=1}^K \|\wb_k(\eta)\|_2^2 > \eta$ for all $\eta\in [0,\eta_{\max}]$.
The sufficiency is true since by Proposition \ref{property: eta and P_eta}(c) there exists at least a value of $\eta$ such that
$\sum_{k=1}^K \|\wb_k(\eta)\|_2^2 \leq \eta$ when {\sf (P1)} is feasible.
To see the necessity part, suppose that there exists an $\eta \in [0,\eta_{\max}]$ such that $\sum_{k=1}^K \|\wb_k(\eta)\|_2^2 < \eta$.
Then according to the fact that $\sum_{k=1}^K \|\wb_k(0)\|_2^2>0$ and the concavity of $y=\sum_{k=1}^K \|\wb_k(\eta)\|_2^2$, there must exist an intersection point between 
$y=\sum_{k=1}^K \|\wb_k(\eta)\|_2^2$  and  $y=\eta$. The existence of such intersection point suggests that   {\sf (P1)} is feasible and thus is a contradiction.
\hfill $\blacksquare$

\section{Proof of Proposition \ref{prop: FD UDD}}\label{appx: proof of FD UDD}
To prove the duality, we separate $\wb_{k}$ into the DL power $p_k^{\rm D}$ and
beamforming direction $\tilde \wb_{k}$, i.e.,
$\wb_{k}=\sqrt{p_k^{\rm D}} \tilde \wb_{k}$ where $||\tilde
\wb_{k}||_2=1$. Then one can write {\sf (P2)} in \eqref{eqn: prob formulation AO dualADC_2} as
\begin{subequations}\label{eqn: prob formulation AO_fixAll_beam_dualADC}
	\begin{align}
	&\min_{\{\|\wb_k\|_2=1\}} \bigg\{	\min_{ \substack{\{p_k^{\rm D}\geq
			0\}, \{p_\ell^{\rm U}\geq 0\}}}~\sum_{k=1}^{K} p_k^{\rm D} \left(
	\tilde \wb_k^H \Bb \tilde \wb_k \right) + \sum_{\ell=1}^L p_\ell^{\rm U} b_\ell
	\\
	&{\rm s.t.}~~
	\frac{p_k^{\rm D}|\hb_{k}^H \tilde \wb_{k}|^2 /\rho_k^{\rm D} }
	{ \sum_{i=1}^K p_i^{\rm D} \tilde \wb_i^H \tilde \Hb_k \tilde \wb_i +
		\hat \sigma_i^2(\pb^{\rm U}) }\geq 1,~\forall k\in \Kc, \label{eqn:
		prob formulation
		AO_fixAll_beamconsD} \\
	&   \frac{p_\ell^{\rm U}| \vb_\ell^H\gb_\ell|^2 /\rho_\ell^{\rm U}}
	{                 \sum_{k=1}^K p_k^{\rm D} \tilde \wb_k^H \LambdaB(\vb_\ell ,\Rb_{\Phib_{0}}) \tilde \wb_k+\sum_{j =1 }^L p_j^{\rm U} \vb_\ell^H \tilde \Gb_j \vb_\ell + \tilde \sigma_z^2 } \geq 1, \notag \\
	&~~~~~~~~~~~~~~~~~~~~~~~~~~~~~~~~~~~~~~~~~~~~~~~~~\forall \ell\in
	\Lc. \label{eqn: prob formulation
		AO_fixAll_beamconsU} \bigg\}
	\end{align} 
\end{subequations}
Since the inner problem of \eqref{eqn: prob formulation AO_fixAll_beam_dualADC} is a linear programming satisfying
the Slater's condition, it has a zero duality gap with its Lagrange dual problem, which can be shown as
\begin{subequations}\label{eqn: DMU dual problem_fixedBeam_dualADC}
	\begin{align}
	\max_{\substack{\lambdab, \mub \succeq \zerob}}  ~ 
	&\lambdab^T\sigmab^2+ \tilde \sigma_z^2 \mub^T \oneb   \label{eqn: DMU dual problem_fixedBeam_dualADC target} \\
	{\rm s.t.}&~~\frac{\lambda_{k}
		|\tilde \wb_k^H\hb_{k}|^2/\rho_{k}^{\rm D} } { \tilde \wb_k^H \Qb(\lambdab,\mub)
		\tilde
		\wb_k } \leq 1,~k\in \Kc, \label{eqn: DMU dual problem_fixedBeam_dualADC VDL}\\
	&\frac{\mu_{\ell}
		|\vb_\ell^H\gb_\ell|^2/\rho_{\ell}^{\rm U}  } {\sum_{j=1}^L \mu_{j}
		\vb_\ell^H \tilde \Gb_j
		\vb_\ell+\sum_{k=1}^{K}\lambda_{k} |f_{\ell k}|^2+b_\ell
	} \leq   1, \ell\in \Lc,
	\label{eqn: DMU dual problem_fixedBeam_dualADC VUL} 
	\end{align}
\end{subequations}
where $\lambdab=[\lambda_1,\ldots,\lambda_K]^T$ and $\mub=[\mu_1,\ldots,\mu_L]^T$ are respectively the dual variables associated with constraints \eqref{eqn:
	prob formulation
	AO_fixAll_beamconsD} and \eqref{eqn: prob formulation
	AO_fixAll_beamconsU}.
Now consider the following problem that is obtained by changing the `$\max$' to `$\min$' and `'$\leq$' to `$\geq$' in \eqref{eqn: DMU dual problem_fixedBeam_dualADC}
\begin{subequations}\label{eqn: DMU dual problem_fixedBeam_dualADC 2}
	\begin{align}
	\min_{\substack{\lambdab, \mub \succeq \zerob}}  ~ 
	&\lambdab^T\sigmab^2+ \tilde \sigma_z^2 \mub^T \oneb   \label{eqn: DMU dual problem_fixedBeam_dualADC target 2} \\
	{\rm s.t.}&~~\frac{\lambda_{k}
		|\tilde \wb_k^H\hb_{k}|^2/\rho_{k}^{\rm D} } { \tilde \wb_k^H \Qb(\lambdab,\mub)
		\tilde
		\wb_k } \geq 1,~k\in \Kc, \label{eqn: DMU dual problem_fixedBeam_dualADC 2 VDL}\\
	&\frac{\mu_{\ell}
		|\vb_\ell^H\gb_\ell|^2/\rho_{\ell}^{\rm U}  } {\sum_{j=1}^L \mu_{j}
		\vb_\ell^H \tilde \Gb_j
		\vb_\ell+\sum_{k=1}^{K}\lambda_{k} |f_{\ell k}|^2+b_\ell
	} \geq   1, \ell\in \Lc.
	\label{eqn: DMU dual problem_fixedBeam_dualADC 2 VUL} 
	\end{align}
\end{subequations}
It is easy to verify that constraints \eqref{eqn: DMU dual problem_fixedBeam_dualADC VDL} and \eqref{eqn: DMU dual problem_fixedBeam_dualADC VUL} of problem \eqref{eqn: DMU dual problem_fixedBeam_dualADC} hold with equality at the optimum. Therefore, problem \eqref{eqn: DMU dual problem_fixedBeam_dualADC 2}  is feasible and 
constraints \eqref{eqn: DMU dual problem_fixedBeam_dualADC 2 VDL} and \eqref{eqn: DMU dual problem_fixedBeam_dualADC 2 VUL} also hold with equality at the optimum.
Based on this fact, it is not difficult to verify that  problems \eqref{eqn: DMU dual problem_fixedBeam_dualADC} and \eqref{eqn: DMU dual problem_fixedBeam_dualADC 2} have the same set of KKT conditions and therefore the two problems achieve the same optimal objective value.
\hfill $\blacksquare$

        \vspace{-0.1cm}
        \footnotesize
        \bibliography{distributed_opt,bf,refs10}
    \end{document}